\documentclass[pre,twocolumn,showpacs,superscriptaddress]{revtex4}

\usepackage{graphics}
\usepackage{amsmath}
\usepackage{amssymb}

\newcommand{\p}{\partial}
\newcommand{\Ai}{\operatorname{Ai}}
\newcommand{\be}{\begin{equation}}
\newcommand{\ee}{\end{equation}}
\newcommand{\bma}{\begin{bmatrix}}
\newcommand{\ema}{\end{bmatrix}}
\begin{document}

\title{Elastic wave propagation in confined granular systems}

\author{Ell\'ak Somfai}
\email{ellak@lorentz.leidenuniv.nl}
\affiliation{Instituut--Lorentz, Universiteit Leiden, PO Box 9506,
        2300 RA Leiden, The Netherlands}

\author{Jean-No\"el Roux}
\affiliation{Laboratoire des Mat\'eriaux et des Structures du G\'enie
  Civil, Institut Navier,
        2 all\'ee Kepler, Cit\'e Descartes, 77420 Champs-sur-Marne, France}

\author{Jacco H. Snoeijer}
\affiliation{Instituut--Lorentz, Universiteit Leiden, PO Box 9506,
        2300 RA Leiden, The Netherlands}
\affiliation{ESPCI, 10 Rue Vauquelin, 75231 Paris Cedex 05,
        France}

\author{Martin van Hecke}
\affiliation{Kamerlingh Onnes Lab, Universiteit Leiden, PO Box 9504,
        2300 RA Leiden, The Netherlands}

\author{Wim van Saarloos}
\affiliation{Instituut--Lorentz, Universiteit Leiden, PO Box 9506,
        2300 RA Leiden, The Netherlands}

\date{\today}

\begin{abstract}
We present numerical simulations of acoustic wave propagation in
confined granular systems consisting of particles interacting with
the three-dimensional Hertz-Mindlin force law. The response to a
short mechanical excitation on one side of the system is found to
be a propagating coherent wavefront followed by random
oscillations made of multiply scattered waves. We find that the
coherent wavefront is insensitive to details of the packing: force
chains do not play an important role in determining this
wavefront. The coherent wave propagates linearly in time, and its
amplitude and width depend as a power law on distance, while its
velocity is roughly compatible with the predictions of macroscopic
elasticity. As there is at present no theory for the broadening
and decay of the coherent wave, we numerically and analytically
study pulse-propagation in a one-dimensional chain of identical
elastic balls. The results for the broadening and decay exponents
of this system differ significantly from those of the random
packings. In all our simulations, the speed of the coherent
wavefront scales with pressure as $p^{1/6}$; we compare this
result with experimental data on various granular systems where
deviations from the $p^{1/6}$ behavior are seen. We briefly
discuss the eigenmodes of the system and effects of damping are
investigated as well.
\end{abstract}

\pacs{45.70.-n, 43.40.+s, 46.40.Cd, 46.65.+g}
% 45.70.-n      Granular systems
% 43.40.+s      Structural acoustics and vibration
% 46.40.Cd      Mechanical wave propagation (including diffraction,
%                 scattering,  and dispersion)
% 46.65.+g      Random phenomena and media

\maketitle

\section{Introduction}

The behavior of granular systems is often strongly influenced by
fluctuations and inhomogeneities on the scale of the granular
particles, which precludes or complicates the derivation of
macroscopic laws from grain-level characteristics. In quasi-static
granular packings the strong fluctuations in the intergrain forces
are a striking example of this heterogeneity \cite{howell99}. The
fact that the grain--grain contacts which support large forces are
usually correlated in a line-like fashion over distances of
several particle diameters leads to so called \emph{force chains}. 
(In static equilibrium a contact transmitting a large force is often balanced
with a single contact on the opposite side of the grain, and this is repeated
on several subsequent grains.) 
These are clearly visible in experiments on two-dimensional (2D)
packings using photoelastic disks \cite{howell99} and in
simulations, even though their precise definition is not agreed on
at present.

More importantly, it is not clear what properties of the granular
media are affected by these force chains or indeed by the broad
distribution of interparticle forces.
It is under debate whether granular media can be described as elastic, and
since a static granular packing is a quenched system, an important issue is at
what length scale such a continuum (elastic) theory would become appropriate
for a granular medium. 

An important quantity that can probe these issues is the
propagation of sound waves. Surprisingly, even the scaling with
pressure of a continuum quantity like the speed of sound is still
a matter of debate---even though clearly the average interparticle
force scales with the pressure, the number of interparticle
contacts also increases with pressure, and it is being debated
whether this might affect the dependence of the speed of sound on
pressure \cite{makse99,velicky02,GO90}. Moreover, one might argue
as follows for an interplay between force chain type correlations
and sound propagation. In a simple 1D array of grains the group
velocity of waves is higher for a more compressed array  because
of the nonlinearity of the force law \cite{coste97}.  Similarly, a
simple calculation shows that in a continuum elastic medium with a
hard block embedded in soft surrounding, the small amplitude
acoustic waves propagate principally in the hard material.  From
these observations one might wonder whether the acoustical waves
that travel through the granular medium first are transmitted
mostly by the strong force chains. If so, one might be able to
extract detailed information about the force chains from the
transmitted acoustic signal.

In this paper we explore numerically and analytically what
ultrasound experiments \cite{ultrasound} might tell about the
microscopic and mesoscopic features of granular media:  can they
tell us anything about the existence of force chains? Can
ultrasound probe on what length scale we can view a granular
medium as an almost homogeneous random medium? Are there generic
features of the propagation of a short pulse which we can uniquely
relate to the random nature of a granular pack? These are some of
the issues we have in mind in this paper.

Using numerical simulations in 2D and 3D, we show
that sound does not predominantly travel along force chains: the
leading part of the wave, which propagates after being excited by
a short pulse at one end of the medium, is better characterized as
a propagating rough front, like what was found in bond-diluted
models~\cite{astrom97}. Similarly to experiments of this type, we
can separate the transmitted acoustic signal into an initial
coherent part and a subsequent random part. The coherent wave
propagates essentially linearly in time, defining a time-of-flight
sound velocity. We have calculated the effective elastic constants
for our packings, which under the assumption that the granular
packing can be viewed as a continuum directly lead to a simple
expression for this velocity. Surprisingly, the observed
time-of-flight velocity of the coherent wave is roughly 40\%
larger than predicted by continuum elasticity.

We also study the scaling of sound velocity and elastic constants
with pressure. For a fixed contact network with Hertzian forces
which stay proportional to the pressure $p$, the elastic constants
scale as $p^{1/3}$ and the sound velocity as $p^{1/6}$.  In the pressure range
studied here we find no discernible deviation from the $p^{1/6}$
behavior for the time of flight velocity; the situation for the
velocities calculated from the effective elastic constants is more
convoluted.

Discrete numerical simulations have been used to evaluate the
elastic moduli of packings of spherical beads before
~\cite{makse99,ohern03,JNR04}, from which continuum sound speeds
can be deduced. However, to our knowledge, wave \emph{propagation}
has never been directly addressed numerically in a granular
material. Measurements of overall elastic properties do not probe
the material on the mesoscopic scale and overlook potentially
interesting properties such as dispersion and attenuation. Hence
the interest and motivation of the present work, which directly
copes with the properties of a traveling ultrasonic pulse in a
model granular packing.

We find that there are various other interesting aspects of the
problem of pulse propagation in granular systems which appear to
have received little attention so far: for disordered systems the
amplitude of the coherent wave decays as a power law as it
propagates, while its width increases linearly. As we are unaware
of systematic studies of these issues so far, we also consider, as
a reference,  the propagation of a sound pulse in 1D chains and in
2D triangular lattices of identical elastic balls. We show that
also here both the amplitude and width of the coherent wave behave
as power law of the distance. We calculate both these exponents
and the wave front analytically, and show that the broadening of
the pulse and the decay of the width is much slower  than in the
2D disordered system. A more detailed experimental and theoretical
study of these aspects might therefore yield an important way to
probe granular media in the future.

This paper is organized as follows. First we review in
Sec.~\ref{sec:experiments} some of the relevant results of
experiments and numerical simulations. Then Sec.~\ref{sec:model}
describes the details of our numerical model: how the packings are
obtained and how the small amplitude oscillations are analyzed. In
Sec.~\ref{sec:results} we present qualitative and quantitative
results of our simulations.  In Sec.~\ref{sec:analytic} we compare
these results to theoretical models of a 1D chain and a triangular
lattice of identical balls.  Finally Sec.~\ref{sec:summary}
concludes the paper.

\section{Elasticity and sound propagation: experimental results
\label{sec:experiments}}

There have been a number of related experiments that considered
sound propagation in granular systems. Because of their
direct relevance we briefly review their main results. It is
useful to keep in mind that, in principle, two types of
experiments can be performed: either one drives the system with
constant frequencies and focuses on spectral properties, or one
drives the system with short pulses, testing propagative features.
Since wave propagation is traditionally described in the framework of
macroscopic linear elasticity, we also briefly evoke some of
the measurements of elastic moduli as obtained in laboratory
experiments.

Liu and Nagel \cite{liu92,liu93,liu94} studied acoustic sound
propagation through an open 3D granular assembly. They prepared a
15--30 grain diameter deep layer of glass beads in an isolated
box. In their setup the top surface was free and subject to
gravity only. The sound source was a vertical extended plate,
embedded within the granular layer, and the detectors were
accelerometers of size comparable to the grains. They identified
three distinct sound velocities. From the response to a short
pulse, the ratio of the source--detector distance and the time of
flight gives $c_\text{tof}=L/T_\text{flight} = 280\pm30$ m/s,
while the dependence of the time of maximum amplitude of the
response on source--detector distance yields
$c_\text{max.resp}=dL/dT_\text{max}= 110\pm 15$ m/s.  In case of
harmonic excitation, the group velocity defined by the frequency
dependence of the phase delay is $c_\text{group} = 2\pi L
d\nu/d\phi = 60 \pm 10$ m/s.  From these incompatible values they
concluded that the granular packing cannot be considered as an
effective medium for sound propagation, as the transmission is
dominated by the strong spatial fluctuations of force networks.  A
consequence of this is the extreme sensitivity on small changes
(e.g., heating one bead by less than a degree \cite{liu94}), and
the $f^{-2}$ power spectrum on response to harmonic excitation.
Many aspects of these experiments have been confirmed by numerical
simulations in a model system of square lattice of non-identical
springs \cite{leibig94} and with Hertzian spheres \cite{melin94}.

Jia and coworkers \cite{jia99,jia01} measured sound propagation through a
confined 3D granular system. They filled a cylinder of radius and length
15--30 grain diameters with glass beads, compacted by horizontal
shaking,
closed off with a piston, and applied an external pressure on the piston. The
sound source was a large flat piezo transducer at the bottom of the cylinder,
and the (variable size) detector was at the top wall.
They measured the response function for a short pulse, and observed that it
can be divided into an initial \emph{coherent} part, insensitive to the
details of the packing, followed by a noisy part, which changed
significantly
from packing to packing.  The ratio of the amplitudes of the two parts of the
signal depended on the relative size of the detector and the grains, and also
on additional damping (e.g., wet grains \cite{jia01}).

Gilles and Coste \cite{gilles03} studied sound propagation in a confined 2D
lattice of steel or nylon beads. They arranged the beads on a triangular
lattice of hexagonal shape, 30 beads on each side, and isotropically
compressed the system.
This resulted in a regular array of bead centers, but irregular
intergrain contacts because of the small polydispersity of the beads.  Both
the sound source and the sensor were in contact with a single bead at opposite
sides of the hexagon.
They also found that the response to a short pulse was an initial coherent
signal, followed by an incoherent part. While the whole signal was
reproducible for a fixed setup, they observed a high correlation factor between
the coherent signals of different packings, and low correlation between the
incoherent parts.

Apart from these recent experiments carried out in the physics
community, it is worth recalling that many interesting
measurements of the elastic or acoustical characteristics of
granular materials are to be found in the soil mechanics and
geotechnical engineering literature (~\cite{TAT01} is a recent review
stressing the need for sophisticated rheological measurements of
soils, including elastic properties).
The elastic properties of granular soils have
been investigated from quasistatic stress-strain dependencies, as
measured with a triaxial~\cite{HI96} or a hollow
cylinder~\cite{GBDC03} apparatus, by ``resonant column''
devices~\cite{CIJ88a,CIJ88b} (which measure the frequency of the
long wavelength eigenmodes of cylindrically shaped samples), and
from sound propagation
velocities~\cite{THHR90,GBDC03,ZENI99,SDH04,TAT01}. One remarkable
result is the consistency of moduli values obtained with various
techniques, provided the applied strain increments are small
enough (\emph{i.e.}, typically, lower than $10^{-5}$). Thus, the
agreement between sound propagation and resonant column results is
checked \emph{e.g.} in~\cite{THHR90}, while~\cite{GBDC03} shows
consistency between sound propagation velocities and static moduli
for very small strain intervals.

In that community, wave velocities are most often deduced
from signal time-of-flight measurements between pairs of specifically
designed, commercially available piezoelectric transducers known as
\emph{bender elements}~\cite{THHR90,JOCO98}
which couple to transverse modes or  \emph{bender-extender elements},
which also excite longitudinal waves~\cite{LIGR01}. The
typical size of such devices is $\sim 1$cm, sand grains diameters
are predominantly in the $100\mu$m range, a typical propagation length
(specimen height or diameter) is 10cm, and confining stresses range
from 50 or 100 kPa to several MPa. Those experiments are thus
comparable to the ones by Jia~\cite{jia01}, the material being however
probed on a somewhat larger length scale. The shape of signals
recorded by the receiver are similar (see \emph{e.g.}~\cite{ZENI99,SDH04}).

The soil mechanics literature on wave propagation is chiefly concerned
with the measurement of macroscopic elastic moduli, with little
interest directed to small scale phenomena. After extracting the
wave velocity from the ``coherent'' part of the signal, using an
appropriate procedure, as discussed \emph{e.g.} in~\cite{VIAT95,JCS96},
the rest is
usually discarded as ``scattering'' or ``near field'' effects.
Most of these experiments are done in sands, rather than assemblies of
spherical balls (see, however, \cite{SDH04}).

To summarize: it appears that granular systems can be considered
as an effective medium for the transmission of a short acoustic
pulse, if probed on sufficiently large scales and pressures, and
if one focuses on the initial part of the response.
This wave front is however followed by a noisy tail, which is
sensitive to packing details, and any quantity or measurement that
is dominated by the noisy part, such as $c_\text{max.resp}$ or
$c_\text{group}$, will not show the effective medium behavior.

Also, probed on smaller scales or possibly at smaller
pressures, the effective medium description appears to become
less accurate, even for the initial coherent wave front. The two
outstanding questions are thus to identify under what conditions
continuum descriptions hold, and if they fail, what other
mechanisms come into play.

\section{Numerical model}
\label{sec:model}
Although we performed numerical simulations on 2D and 3D granular
packings, this paper mainly contains 2D results. In our setup
(most similar to the experiments of Jia \emph{et al.}
\cite{jia99,jia01}) we have a rectangular box containing confined
spheres under pressure. We send acoustic waves through one side of
the box and detect force variations at the opposite side.

First we prepare a static configuration of grains. We start with
a rectangular box filled with a loose granular gas. For the 2D
simulations the spheres have polydispersity to avoid
crystallization: the diameters are uniformly distributed between
0.8 and 1.2 times their average. The bottom of the box is a solid
wall, we have periodic boundary conditions on the sides, and the
top is a movable piston.  We apply a fixed force on the (massive)
piston, introduce a Hertz-Mindlin force law, friction included, with some
dissipation for the intergrain collisions (see appendix \ref{Hertzapp}), and
let the system evolve until all motion stops.  Our packings are
considered to have converged to mechanical equilibrium when all
grains have acceleration less than $10^{-10}$ in our reduced
units (defined immediately below).  At this point we have a static granular
system under external pressure.

In the rest of the paper we use the following conventions.  Our
unit of length is the average grain diameter.  The unit of mass is
set by asserting that the material of the grains has unit density.
We set the individual grain's modified Young modulus $E^*=1$, which becomes
the pressure unit (see appendix~\ref{Hertzapp}). Since the grains are
always 3D spheres, we measure pressure even in the 2D case as ``3D
pressure'': force divided by area, where in 2D the area is length
of box side times grain diameter.  The speed of sound of the
pressure waves inside the grains becomes unity (for zero Poisson
ratio).  This sets our unit of time, which is about an order of
magnitude shorter than the time scale of typical granular
vibrations.

Most results are obtained on series of (approximately) square 2D
samples containing 600 spheres (their centers being confined to a
plane), prepared with friction coefficient $\mu=0.5$ and Poisson
ratio $\nu=0$ (see appendix~\ref{Hertzapp}). The confining stress
$p$ that is controlled in the preparation procedure, and referred
to as the pressure throughout this article, is actually the
(principal) stress component $\sigma _2$ (or $\sigma _{22}$ in a
system of axes for which coordinate labeled ``2'' varies
orthogonally to the top and bottom solid walls). To check for the
influence of $p$ on the results, we prepared samples under
pressures $p=10^{-7}$, $10^{-6}$, $10^{-5}$ and $10^{-4}$ (30
samples for each value). To gain statistical accuracy we also
prepared an additional series of 1000 samples at $p=10^{-4}$. If
the particles are glass beads this corresponds to $7\mbox{kPa} \le
p \le 7\mbox{MPa} $, an interval containing the pressure range
within which solid granular packings are usually probed in static
or sound propagation experiments. It is worth pointing out that
the assembling procedure is repeated for each value of the
pressure. It results in a specific anisotropic equilibrium state
of the granular assembly, as characterized in
section~\ref{subsec:static}.

To check the robustness of qualitative results on sound
propagation, we also studied  a few 3D samples, obtained similarly
to 2D samples by a compaction of a monodisperse granular gas with a
piston compressing in the $z$ direction to
$p=\sigma_{33}=10^{-4}$, and periodic in $x$ and $y$ directions.
 
In most of the work below, with the exception of section \ref{subsec:damping},
we study small amplitude oscillations.  For this we can use a system
linearized around its equilibrium: in the static packing we replace the
intergrain contacts with linear springs, with stiffness obtained from the
differential stiffnesses (essentially $ d F_\text{t} / dt $ and
$d F_\text{n} / dn$)
  of the individual Hertz-Mindlin contacts (see appendix \ref{Hertzapp}).
   The equation
of motion becomes
\begin{equation}
\mathsf{M}\ddot{\mathbf{u}} = - \mathsf{D}\mathbf{u} \,,
\end{equation}
where for $N$ grains the vector $\mathbf{u}$ contains the $3N$ coordinates and
angles (in 2D) of the particle centers, $\mathsf{M}$ is a diagonal matrix
containing grain masses and moments of inertia, and $\mathsf{D}$ is the
dynamical matrix containing information about contact stiffnesses and
the network topology.

Then we solve the eigenproblem of the linear spring system and write the
oscillation of the grains as the superposition
\begin{equation}
\mathbf{u}(t) = \sum_n a_n \hat{\mathbf{u}}^{(n)} \sin(\omega_n t) \,.
\end{equation}
The amplitudes $a_n$ are obtained by the projection of the initial condition
onto the eigenmodes.  When the force on the top wall is calculated, we
calculate the coupling $b_n$ of the given mode with the wall, so the force
is
\begin{equation}
\label{eq:Ftop}
F_\text{top} = \sum_n a_n b_n \sin(\omega_n t) \,.
\end{equation}

Typically we will look at the transmission of a short pulse
through the granular packing.  We send in a delta-pulse, corresponding to
wide band excitation.  This corresponds to the following initial condition:
at $t=0$ the grains in contact with the bottom wall have a velocity
proportional to the stiffness of the contact with the wall.
This is equivalent to an infinitesimally short square pulse: raising the
bottom wall for an infinitesimally short time and lowering it back. The
amplitude of the pulse does not matter as all the calculations are linear
(except section \ref{subsec:damping}). The quantity we
wish to study are the resulting force variations on the top wall
(the ``signal''), defined as the force exerted by the vibrating
grains on the top wall minus the static equilibrium force:
\begin{equation}
F_\text{sig}(t)=F_\text{top}(t)-F_\text{top}(0^-)~.
\end{equation}

\section{Numerical results}
\label{sec:results}

In this section we will present the results of extensive numerical
simulations of sound propagation in granular media. We first
characterize the geometric state of the packing and measure its
static properties, including the tensor of elastic moduli. Then we
report wave propagation simulations, showing that in our system,
like in the experiments of Jia \emph{et al.} \cite{jia99,jia01},
$F_\text{sig}$ is composed of a coherent initial peak and an
incoherent disordered tail. We find that force lines are fairly
irrelevant for the pulse propagation, and we will show that the
decay and spreading of the initial peak follow
packing-detail-independent scaling laws. Furthermore we study the
pressure dependence of the transmission velocity of the initial
pulse, which is compared to the one deduced from static elastic
properties. We close this section by a short discussion of the
spectrum and eigenmodes that determine the sound propagation, and
briefly study the effect of dissipation.

\subsection{Statics \label{subsec:static}}

\begin{table*}[bht]
\centering
\begin{tabular}{|c|cccccccccc|}  %\cline{1-5}
\hline
$p$&$\Phi_2$& $\Phi_3$ & $\Phi_3^*$ & $z^*$& $f_0$ (\%)& $\langle N/p
\rangle$& $Z(2)$ & $Z(3)$ & $F_2$&$F_2^S$\\
\hline
 $10^{-7}$
& $0.818\pm 0.005$
& $0.561\pm 0.005$
& $0.48\pm 0.01$
& $3.18\pm 0.03$
& $15\pm 2$
& $1.31$ &$1.49$& $2.81$
& $0.527$&$0.557$\\
\hline
 $10^{-6}$
& $0.814\pm 0.004$
& $0.558\pm 0.005$
& $0.49\pm 0.01$
& $3.23\pm 0.04$
& $14\pm 2$
& $1.27$& $1.50$& $2.83$
& $0.524$&$0.560$\\
\hline
 $10^{-5}$
& $0.809\pm 0.004$
& $0.554\pm 0.005$
& $0.50\pm 0.01$
& $3.31\pm 0.03$
& $11\pm 2$
&$1.22$& $1.49$& $2.77$
& $0.523$&$0.558$\\
\hline
 $10^{-4}$
& $0.815\pm 0.005$
& $0.558\pm 0.005$
& $0.52\pm 0.01$
& $3.48\pm 0.03$
& $7.4\pm 1.5$
&$1.06$& $1.51$& $2.90$
& $0.524$&$0.578$\\
\hline
\end{tabular}
\caption{Static data, corrected for boundary effects: solid fractions,
  coordination number $z^*$, proportion of rattlers $f_0$, average
  normal force, reduced moments (defined
  in~\eqref{eqn:defZ}) of the force distribution, and fabric parameters
  $F_2$ (all contacts) and $F_2^S$ (strong contacts) defined
  in~\eqref{eqn:defF2}. Starred quantities are evaluated on
  discarding rattlers. Stated error bars correspond, here as in all
  subsequent tables, to one standard deviation on each side of the mean.}
\label{tab:static}
\end{table*}

We first characterize the system by its geometric and static
properties, a necessary step as sound propagation is sensitive to the
internal state of a granular packing. Our preparation procedure yields
values given in table~\ref{tab:static} for solid
fractions $\Phi_2$ (2D definition) and  $\Phi_3$ (3D definition, in a
one-diameter thick layer) and proportion of
rattlers (particles that transmit no force) $f_0$. For sound
propagation, the relevant mass density, with our choice of units, is
equal to the 3D solid fraction $\Phi_3^*$ of non-rattler grains.
Values for those parameters are given on accounting for boundary
effects: measurements exclude top and bottom layers of thickness $l$,
and we checked that results became $l$-independent for $l\ge 3$.
Table~\ref{tab:static} also gives the bulk values (corrected for wall
effects) of the mechanical coordination number $z^*$ (\emph{i.e.,} the
average number of force-carrying contacts per non-rattler particle),
the average normal contact force $N$, normalized by the pressure, and
some reduced moments of the distribution of normal contact forces,
defined as:
\be
Z(\alpha) = \frac{\langle N^\alpha \rangle}{\langle N \rangle^\alpha}.
\label{eqn:defZ}
\ee
$Z$ values are characteristic of the shape of the force distribution
(the wider the distribution, the farther from 1 $Z(\alpha)$ for any
$\alpha \ne 1$).

The renewal of compression procedure from a granular gas at each
value of $p$ is responsible for the apparent absence of systematic
density increases over the range of pressure studied here.

As observed in other studies (see,
\emph{e.g.,}~\cite{makse00,JNR04}), a direct compression of a
granular gas with friction yields rather loose samples, with a low
coordination number. The minimum value of $z^*$ deduced from the
condition that the number of unknown contact forces should be at
least equal to the number of equilibrium equations for active
particles is 3 in 2D, and the values given in
table~\ref{tab:static} are only barely larger for the smaller
values of $p$. This means that our samples have a low degree of
force indeterminacy at $p=10^{-7}$. Complete absence of force
indeterminacy is obtained with frictionless beads in the limit of
zero pressure~\cite{isostatic1,isostatic2,pzeropreprint,JNR2000}.
With frictional contacts, it appears that some force indeterminacy
persists even in the limit of zero pressure
\cite{makse00,SEGHL02,pzeropreprint}.

The fraction of rattlers is very large (up to $15$\%), when $z^*$ is close to
3, while the number of rattlers fastly decrease as $z^*$ increases with $p$.
We also computed the state of stress in the samples: due to the
assembling procedure, the principal directions are horizontal
(coordinate $x$, principal value $\sigma_1$)
and vertical (coordinate $y$, principal value $\sigma_2$),
with a larger vertical stress, $\sigma_1 /\sigma_2 <1$.
This ratio (table~\ref{tab:moduli}) stays constant, within
statistical error bars, as the controlled confining stress $\sigma_2=p$
is varied, except at the highest pressure $p=10^{-4}$.
(Stresses are readily evaluated with the usual formula
$$\sigma _{\alpha \beta} = \frac{1}{V} \sum _{i<j} F_{ij}^{\alpha}
r_{ij}^{\beta},$$
where $F_{ij}^{\alpha}$ is the $\alpha$ component of the force exerted on
particle $i$ by particle $j$, and the vector ${\bf r}_{ij}$, pointing
from the center of particle $i$ to the center of $j$, should account for
periodic boundary conditions and involve ``nearest image'' neighbors).

\begin{figure*}
\resizebox{!}{0.68\columnwidth}{\includegraphics*{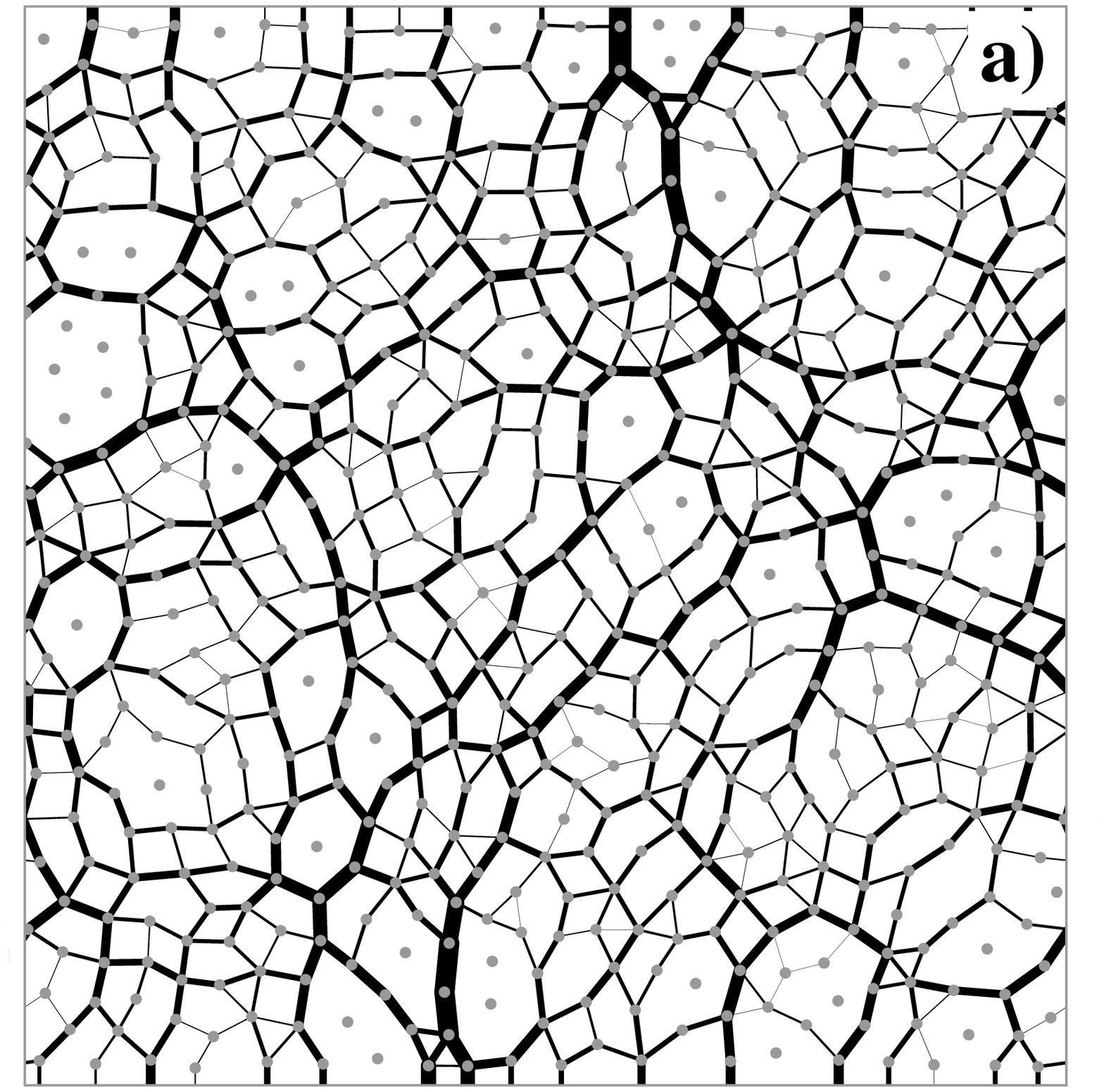}}
\hspace{2mm}
\resizebox{!}{0.68\columnwidth}{\includegraphics*{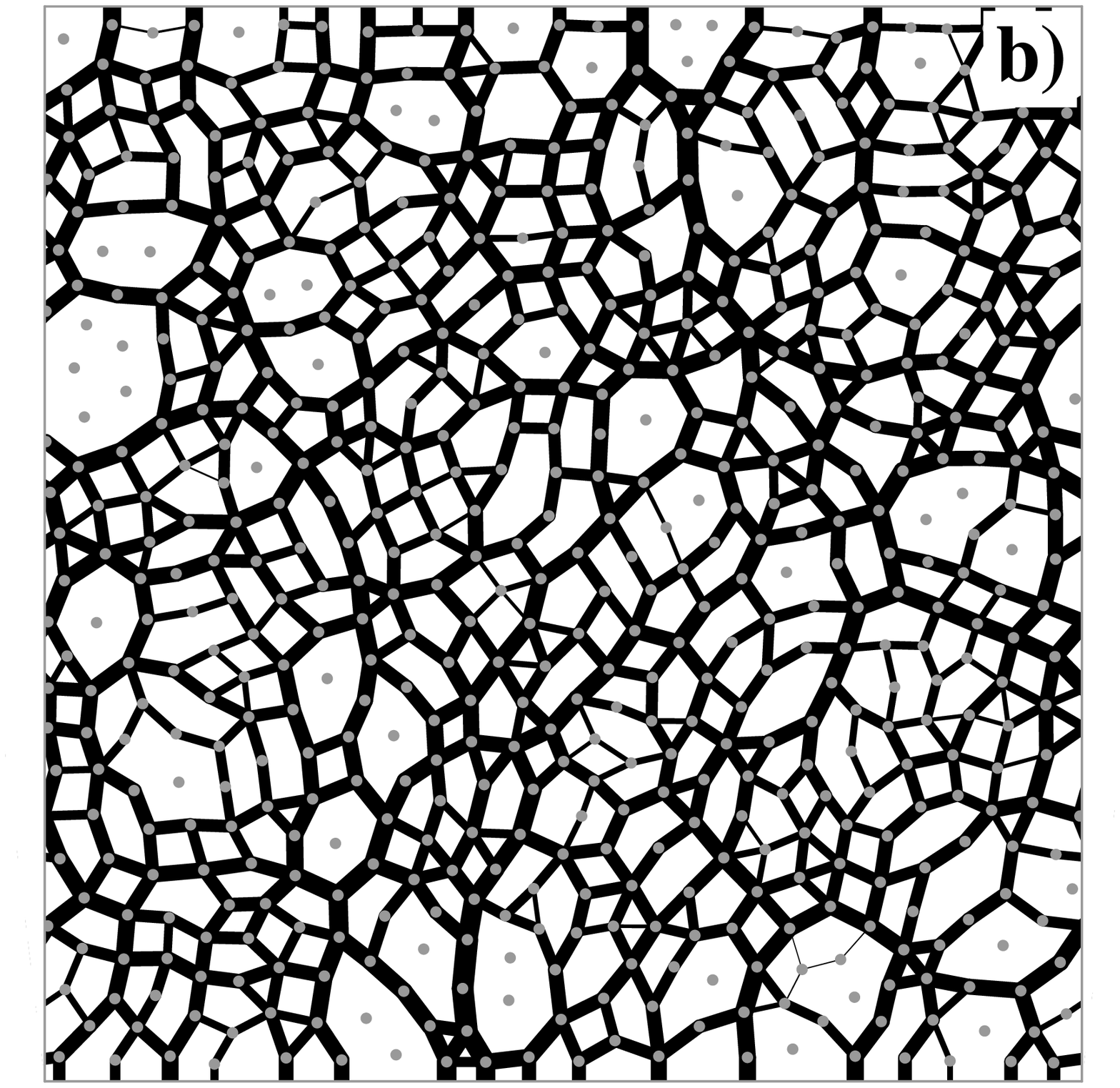}}
\hspace{2mm}
\resizebox{!}{0.68\columnwidth}{\includegraphics*{w23-600.forcehist-lb.eps}}
\caption{\label{fig:static}
  \textbf{(a)} Snapshot of a force network at pressure $p=10^{-4}$.
 Line widths are proportional to the force
between the grains, grain centers are plotted with gray dots.  Only the normal
forces are plotted, the tangential (frictional) are not.
  \textbf{(b)} The stiffness network of the same configuration.  Line widths
are proportional to the stiffness $dF_{\text{n}}/dn$ of the contact (normal
part).  While the
force network shows considerable spatial fluctuations, the stiffness network
is much more homogeneous.
  \textbf{(c)} Histogram of the normal contact forces and contact stiffnesses
of 1000 configurations.  The area under the two curves are the same.  The plot
shows that the stiffness is more narrowly distributed, a feature
discussed in more detail in Sec. \ref{sec:qualitative}.
}
\end{figure*}

\begin{figure*}
\resizebox{1.4\columnwidth}{!}{\includegraphics*{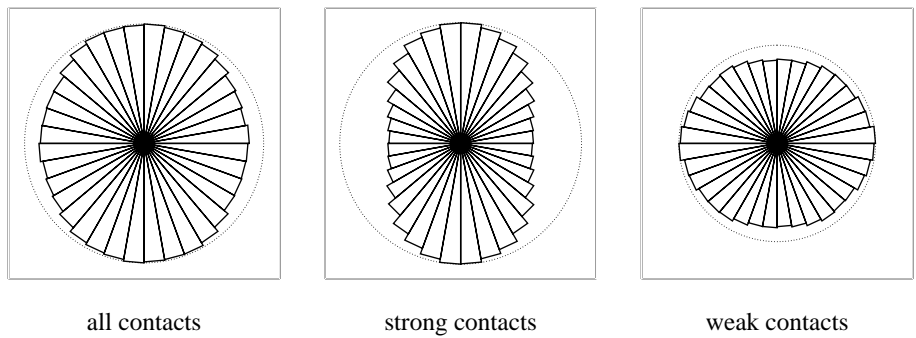}}
\caption{\label{fig:histogram}
  Histogram of contact directions at pressure $p=10^{-4}$.
The diagrams show the average of 1000 independent configurations, 600 grains
each, and only bulk contacts are counted (at least 3 particle diameters away
from walls).  The strong contacts (contact force larger than the average) show
significant anisotropy: vertical directions are favored.  The weak forces are
much more isotropic, although slightly more of them are horizontal than
vertical (as described in~\cite{radjai98}). Recall that the piston compressed the grains in vertical direction.
}
\end{figure*}

The force
network of a typical granular configuration of 600 grains prepared at pressure
$p=10^{-4}$ and friction coefficient $\mu=0.5$ is shown on
Fig.~\ref{fig:static}(a).  We show the stiffness network on
Fig.~\ref{fig:static}(b) and the histogram of forces and stiffnesses on
Fig.~\ref{fig:static}(c), to which we will come back shortly. It is
interesting to note that the shape of the force distribution changes very
little in the investigated pressure range: values of reduced moments
$Z$, listed in table~\ref{tab:static}, hardly change with pressure.
In ref.~\cite{makse00}, a ``participation number'' $\Pi$ was defined,
as an indicator of the width of the force distribution, or the
``degree of localization'' of stresses on force chains. In fact, one would
have $\Pi = 1/Z(2)$ if we had defined $Z$ with the magnitude of the
total contact force instead of its sole normal component. Makse
\emph{et al.}~\cite{makse00} found that $\Pi$ increases
linearly with $\ln p$, until it saturates at $p\ge 10^{-5}$. Our
contradictory observation of a nearly constant $Z(2)$ is likely due
to our compressing the system anew from a granular gas at each $p$,
instead of quasistatically increasing $p$ in a previously assembled
solid sample.

The anisotropy of the contact network (which carries anisotropic stresses) is
apparent in the histogram of contact angles and the force histograms.
Figure~\ref{fig:histogram} shows the  histogram of contact
directions.  We also plot the contribution from strong (contact
force larger than the average) and weak (force smaller than average)
contacts.  Due to the
assembly procedure there is an anistropic stress field, resulting in a
bias of the strong contacts towards the vertical direction. Let us recall
that this is also the principal direction corresponding to the largest
eigenvalue of the stress tensor. As a direct consequence of the force
anisotropy,  the small forces are biased in the opposite  direction,
although this effect is less pronounced~\cite{radjai98}.
One way to quantitatively assess the
importance of this anisotropy is to compute the fabric parameter
\be
F_2 = \langle n_y ^2 \rangle,
\label{eqn:defF2}
\ee
where $n_y$ is the vertical coordinate of the unit normal vector and
the average runs over the contacts. The departure of $F_2$ from its
isotropic value $1/2$ measures the anisotropy. Table~\ref{tab:static} gives
$F_2$ for the different pressure levels, along with parameter $F_2^S$,
obtained on counting only the contacts that carry larger than average
forces. Once again, the level of anisotropy does not depend on
pressure, except for a slight difference at $p=10^{-4}$, in which case
it is a little larger (consistently with the larger stress anisotropy).

If, on long length scales, the material can be considered as an
ordinary homogeneous 2D material with two orthogonal symmetry axes,
the granular packing has 4 independent macroscopic elastic moduli, which
relate the 3 independent coordinates of the symmetric stress
tensor $\sigma _{ij}$ to the 3 coordinates of the symmetric strain
tensor
$\epsilon _{ij}$ as \be \bma  \sigma_{11} \\ \sigma_{22} \\
\sigma_{12}\ema = \bma C_{11} & C_{12} & 0 \\ C_{12} &  C_{22} & 0
\\ 0 & 0 & 2C_{33} \\ \ema \bma  \epsilon_{11} \\ \epsilon_{22} \\
\epsilon_{12}\ema . \label{eqn:moduli} \ee In
Eqn.~\ref{eqn:moduli}, indices 1 and 2 correspond to the
horizontal (periodic) and vertical (along which the normal stress
is controlled) directions on the figures. Counting positively
shrinking deformations and compressive stresses, strain components
should be related to the displacement field ${\bf u}$ as
$$
\epsilon _{ij} = -\frac{1}{2} \left(\frac{\p u_i}{\p x_j}
+\frac{\p u_j}{\p x_i} \right).
$$
Principal stress ratios $\sigma_1/\sigma_2$ and
apparent values of the 4 moduli introduced in Eqn.~\ref{eqn:moduli},
obtained on imposing global strains on the rectangular cell containing
the samples, are given in table~\ref{tab:moduli}.
\begin{table*}[thb]
\centering
\begin{tabular}{|c|ccccc|}  %\cline{1-5}
\hline
Pressure&$\sigma_1/\sigma_2$&$C_{11}$ & $C_{12}$ & $C_{22}$ & $C_{33}$\\
\hline
$10^{-7}$&$0.79\pm 0.06$&$(9.9 \pm 0.9)\times 10^{-4}$&$(8.1 \pm 0.4)\times
10^{-4}$&$(13.9 \pm 0.8)\times 10^{-4}$&$(1.2 \pm 0.4)\times 10^{-4}$\\
\hline
$10^{-6}$&$0.79\pm 0.06 $&$(2.3 \pm 0.2)\times 10^{-3}$&$ (1.7 \pm 0.08)\times
10^{-3}$&$(3.2 \pm 0.2)\times 10^{-3}$&$(0.35 \pm 0.07)\times 10^{-3}$\\
\hline
$10^{-5}$&$0.78\pm 0.06 $&$(5.4 \pm 0.5)\times 10^{-3}$&$ (3.5 \pm 0.2)\times
10^{-3}$&$(7.4 \pm 0.3)\times 10^{-3}$&$(1.1 \pm 0.2)\times 10^{-3}$\\
\hline
$10^{-4}$&$0.71\pm 0.04$&$(1.35 \pm 0.09)\times 10^{-2}$&$ (6.8\pm 0.4)\times
10^{-3}$&$(1.88 \pm 0.08)\times 10^{-2}$&$(4.0 \pm 0.5)\times 10^{-3}$\\
\hline
\end{tabular}
\caption{Stress ratio and elastic moduli, as defined
  in~\eqref{eqn:moduli}, for the 4 investigated pressures with $\mu = 0.5$ and $\nu=0$.}
\label{tab:moduli}
%\normalsize
\end{table*}
To check for possible length scale effects on static elasticity (in
view of the small system size), samples were submitted to
inhomogeneous force fields, or to various local conditions on
displacements. The result of these computer experiments, described in
Appendix~\ref{app:elastic_local}, is that constants $C_{22}$ and
$C_{33}$ are already quite well defined on the (modest) scale of
the 600 sphere samples (typically $24\times 25$). The assumption that
non-uniform stress
and strain fields, which vary on the scale of a fraction of the sample
size, are related by~\eqref{eqn:moduli}, with the elastic constants of
table~\ref{tab:moduli}, predicts results which approximately agree
with numerical tests on our discrete packings, in spite of their
moderate size. This agreement is better for the longitudinal constant
$C_{22}$ than for the shear modulus $C_{33}$, and improves on
increasing $p$.

The conclusion that macroscopic elasticity applies even at moderate
length scales is in agreement with the results by Goldhirsch and 
Goldenberg on homogeneously forced disordered packings
\cite{their_other_paper}.
However, when probing the response to {\em localized} forces
(which perturb the system inhomogenously even in the elastic limit)
those authors identified a larger length scale of about 100 diameters
in order to recover macroscopic elasticity \cite{GOGO02}; these
differences might also be due to the fact that their study concerned 
frictionless quasi-ordered systems.  We should keep in mind also that 
Goldhirsch and Goldenberg looked at the full spatial dependence of the
elastic response, while we extracted only global elastic quantities.
One may also note (see,
\emph{e.g.,}~\cite{RC02}) that constitutive laws are obtained with
numerical simulations of disordered granular samples in the
quasistatic regime with relatively small finite size effects when
the number of particles is above 1000, and that the level of
uncertainty and fluctuations is further reduced on investigating
the response to small perturbations on a fixed contact network.
Tanguy \emph{et al.}~\cite{TWLB02} studied the finite-size effects
on the elastic properties of 2D Lennard-Jones systems at zero
temperature. While large sample sizes ($N\sim 10000$ particles)
were necessary for the low-frequency eigenmodes to resemble the
macroscopic predictions (an issue we shall return to in the sequel
to this paper) they did observe apparent elastic constants, as
measured on globally deforming the sample, to converge quickly
(for $N\sim 100$) to their macroscopic limit. The observation of
macroscopic elastic behavior (with some limited accuracy) in an
assembly of 600 particles is not really surprising in this context.

Assuming macroscopic elasticity to hold in our samples, elastic
constants $C_{22}$ and $C_{33}$ determine velocities of
longitudinal ($c_\ell$) and transverse  ($c_\text{t}$) waves propagating in
direction 2 (normal to top and bottom walls), as
\be c_\ell =
\sqrt{\frac{C_{22}}{\Phi_3^*}}\ \mbox{ and } c_\text{t} =
\sqrt{\frac{C_{33}}{\Phi_3^*}}. \label{eqn:vel_modu} \ee
It should be pointed that the different elastic constants do not exhibit the
same scaling with the pressure. Most notably, the ratio of shear
modulus $C_{33}$ to longitudinal modulus $C_{22}$ (or the ratio of
the corresponding wave velocities, according
to~\eqref{eqn:vel_modu}) steadily increases with $p$. We shall
return to this issue, and compare different predictions for sound
velocities and their pressure dependence, in
section~\ref{subsec:pressure} and in the discussion.

\subsection{Wave propagation: qualitative observations}\label{sec:qualitative}
\begin{figure}
\resizebox{0.9\columnwidth}{!}{\includegraphics*{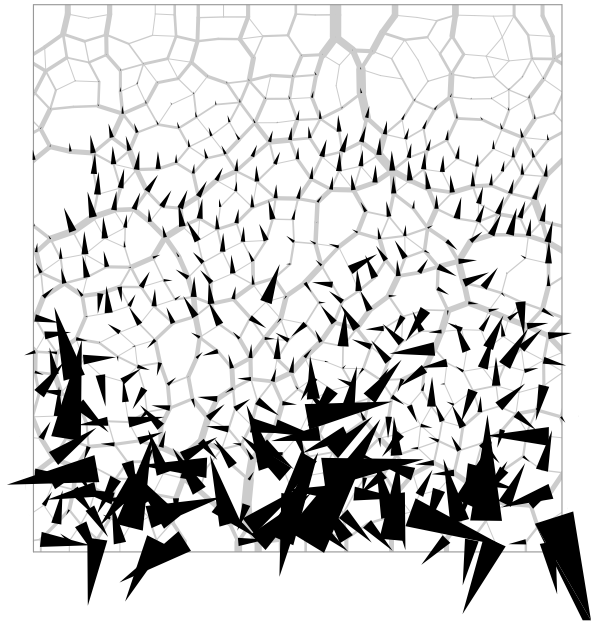}}
\caption{\label{fig:oscill}
  Snapshot of the oscillations.  The length of the arrows show the (magnified)
displacement of the grains from their equilibrium position.  One can see the
localized large amplitude oscillations of the grains near the bottom wall
(source), and a smaller amplitude homogeneous wave traveling towards the top
wall (detector).  At the time of the snapshot, $t=80$, the wave almost reached
the top wall.
}
\end{figure}

We now turn our attention to the pulse propagation. Before
studying $F_\text{sig}$, we will discuss here an example of the spatial
structure of the propagating pulse. Our first observation is that
acoustical waves do not correlate in any obvious way with the
existence of force chain-like configurations. A snapshot of the
grain oscillations shortly after the system is ``kicked'' is shown
on Fig.~\ref{fig:oscill}.

\begin{figure}[!]
\resizebox{0.9\columnwidth}{!}{\includegraphics*{w23-600-signal-lb.eps}}
\medskip

\resizebox{0.9\columnwidth}{!}{\includegraphics*{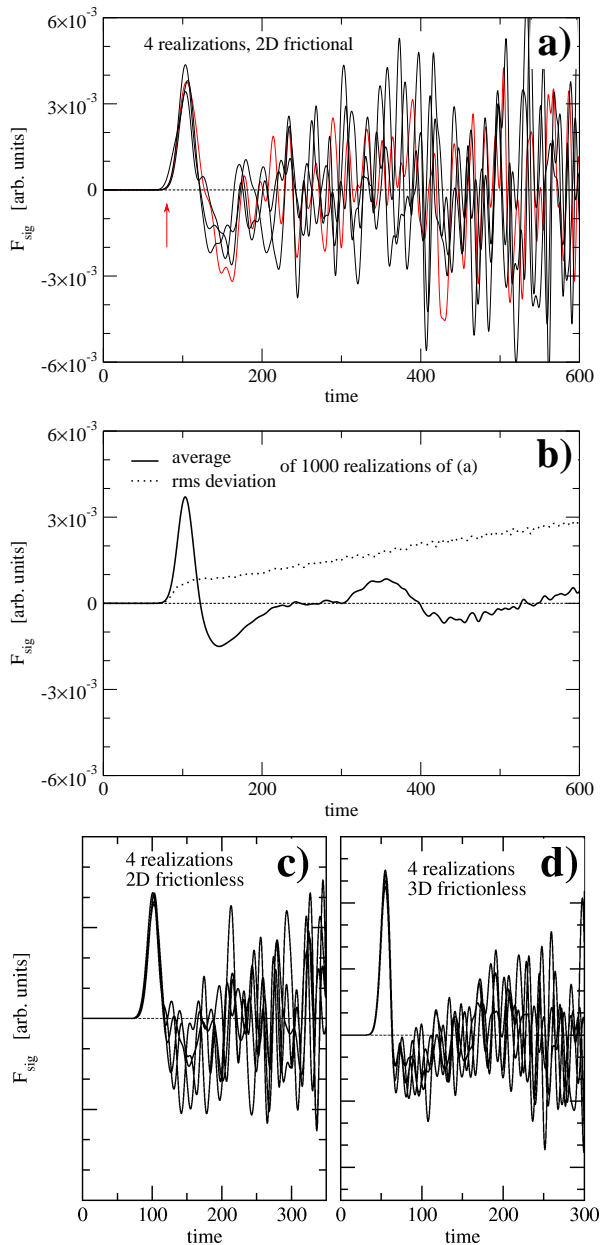}}
\caption{\label{fig:signal}
  (Color online) The signal $F_\text{sig}$,
  the extra force exerted by moving grains on the top wall.
  \textbf{(a)} Four independent 2D frictional configurations are shown, the
signal corresponding to the packing of Fig.~\ref{fig:static} is plotted in
red.  The arrow indicates the time of the snapshot of Fig.~\ref{fig:oscill}.
  \textbf{(b)} The ensemble average and root-mean-square deviation of 1000
independent configurations.  The first cycle of the oscillations is almost the
same on all configurations (we call this the coherent part of the signal),
while the following part is very much configuration dependent.  The ensemble
average only contains the coherent part plus some weak broadened sign of
multiple reflections on the top and bottom walls.
  \textbf{(c)} 2D frictionless and \textbf{(d)} 3D frictionless systems exhibit
similar behavior.
  Time (as well as length in subsequent figures) is denoted in dimensionless
units, see Sec.~\ref{sec:model} for details.
}
\end{figure}

This figure clearly shows that the naive idea that the acoustic
waves would follow the strongest granular force chains is false.
Instead, one can see the propagation of a rough wave front.  One
reason we can immediately point to is that even though the forces
of the intergrain contacts exhibit a strong spatial fluctuation,
the stiffness is much more homogeneous, see
Fig.~\ref{fig:static}(b). This can be understood simply as
follows. Consider a force law which in scaled units reads
$F_\text{n}=n^\beta$, where $n$ is the normal deformation. The
stiffness $s$ is then simply given by $dF_n/dn=\beta n^{\beta-1}$.
Clearly, for $\beta=1$ (corresponding to the 2D Hertzian force
law), all the stiffnesses values are the same. For the
Hertz-Mindlin law, $\beta=3/2$, and we find that the stiffnesses
are proportional to the cubic root of the contact forces, leading
to the rather homogeneous stiffness network shown in
Fig.~\ref{fig:static}(b). So if we compare two links with forces
differing by a factor 8, the corresponding stiffnesses only differ
by a factor 2, and the sound speed---proportional to the square
root of the stiffness---differ only by a factor of $\sqrt{2}$.
Even though the contact forces follow a wide distribution, the
stiffness-distribution is strongly peaked [see
Fig.~\ref{fig:static}(c)]. Although this is a rather trivial
observation for Hertzian contacts, we are not aware of its being
explicitly mentioned in the literature.

An additional reason for the weak effect of force chains on the
sound propagation may be that the disorder of the grains is
significant: on a force chain with weak side links the oscillation
quickly spreads into its neighborhood, resulting in a more
homogeneous base of the oscillations.  Anyway, the conclusion we
can draw here is that the force chains are not relevant for the
evolution of the initial wavefront.

\subsection{The coherent wavefront}\label{subseq_coh}

Let us now study the experimentally accessible
signal $F_\text{sig}$. The time dependence of this signal is shown
on Fig.~\ref{fig:signal}(a). Clearly $F_\text{sig}$ can be thought
to be composed of an initial peak followed by a long incoherent
tail. One can see that for configurations that are similar in
overall geometry but statistically independent, the initial first
cycle of the signal is very similar, but the following part is
strongly configuration dependent. The time dependence of the
signal is very reminiscent of the  traces measured by Jia {\em et
al.}~\cite{jia99} in their ultrasound experiments. Following the
nomenclature introduced by these authors,  we call the first part
of the signal the \emph{coherent} part.  In the ensemble average
only the coherent part of the signal shows up (plus its later
weaker echoes) [see Fig.~\ref{fig:signal}(b)].  The random part of
the signal contributes to the root-mean-square deviation. We also
found that qualitatively, $F_\text{sig}$ is very similar for 2D
frictional, 2D frictionless and 3D frictionless systems, see
Fig.~\ref{fig:signal}(a,c,d).

\begin{figure}
\resizebox{0.9\columnwidth}{!}{\includegraphics*{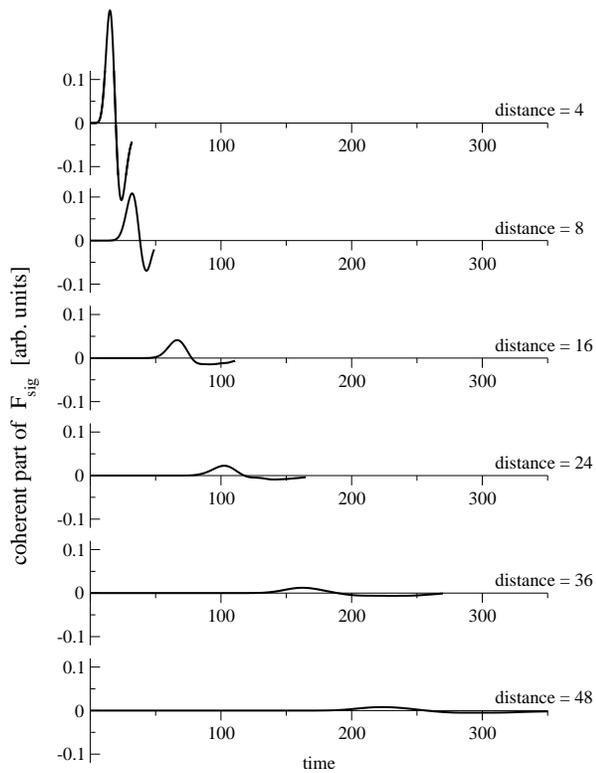}}
\caption{\label{fig:coherent}
  The coherent part of the signal in containers of varying height
(source--detector distance). For taller containers the signal arrives later
with decreased amplitude and increased width. These are quantitatively
analyzed in the next few figures.
}
\end{figure}

We will now focus on the initial peak of $F_\text{sig}$, and
determine for this coherent wavefront its propagation velocity,
and the time evolution of its shape. We have only measured the
time dependence of the signal at a \emph{fixed distance}, and a
qualitative picture of the evolution of the coherent wave can be
extracted from a sequence of measurements at varying
source--detector distance.  This is shown on
Fig.~\ref{fig:coherent}: during the propagation of the signal (as
it arrives later at longer distances), the coherent part's
amplitude decreases, and its width increases.

\begin{figure}
\resizebox{0.9\columnwidth}{!}{\includegraphics*{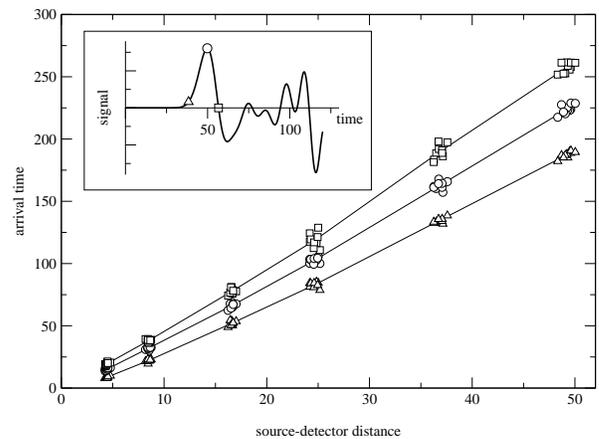}}
\caption{\label{fig:arrival}
  The arrival time of the coherent part of the signal as a function of the
source-detector distance.  The \textbf{inset} shows the definition of the
symbols: leading edge at 10\% of the first peak height ($\triangle$), the
first peak ($\bigcirc$), and the first zero crossing of the signal ($\Box$).
All three characteristic points of the signal have a linear time--distance
relation.  The \emph{slope} of the time--distance plot of the leading edge
defines a time-of-flight velocity: $c_\text{tof}= 0.25$.
}
\end{figure}

\begin{figure}
\resizebox{0.9\columnwidth}{!}{\includegraphics*{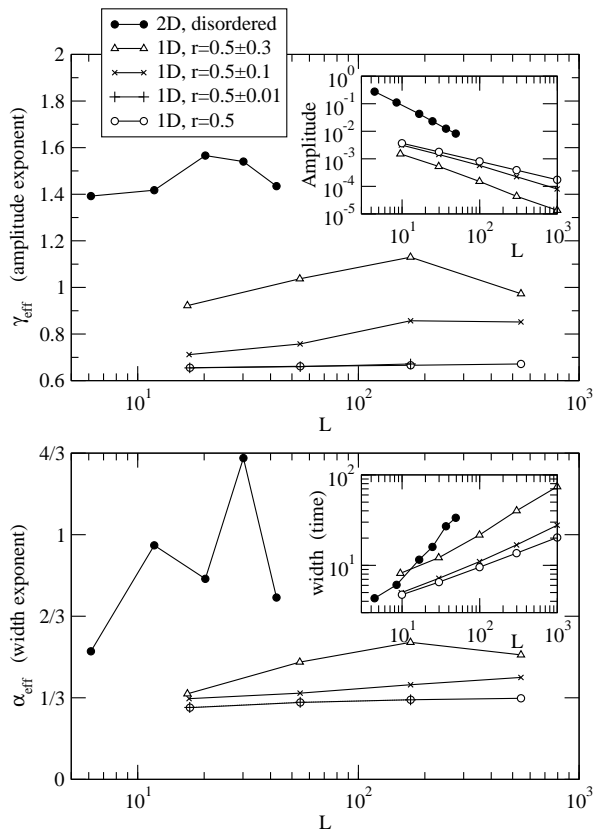}}
\caption{\label{fig:amplitude-width}
  The scaling of the amplitude and width of the coherent part of the signal
with the source--detector distance $L$.
  \textbf{Upper panel:} the amplitude follows roughly $A\sim L^{-\gamma}$
(inset).
In the main panel we plot the effective value of the exponent:
$\gamma_\text{eff}=d\log A/d\log L$.  The symbols are 2D disordered (full
circles),
1D chain of identical balls (open circles) and 1D chain of polydisperse balls
(other symbols, with varying polydispersity).
  \textbf{Lower panel:} the width of the coherent part of the signal increases
with distance: width $\sim L^\alpha$ (inset).  In the main panel here also the
effective exponent $\alpha_\text{eff}=d\log W/d\log L$ is plotted.
}
\end{figure}

Now we look at $F_\text{sig}$ quantitatively. As shown in the
inset of Fig.~\ref{fig:arrival}, we characterize the coherent peak
by three points: its peak location, its first 10\% of peak value
and its first zero crossing. In Fig.~\ref{fig:arrival} we show,
for various source-detector distances, the times at which these
three characteristics can be observed at the detector. In
reasonable approximation, the time of flight depends linearly on
the source--detector distance, although the upward curving of the
data suggest that for small systems the propagation velocity
appears larger than for large systems. We define the time of
flight velocity, $c_\text{tof}$, by measuring the difference of the 
arrival times (at 10\% of the peak's level) for
source-detector distances of 8 and 25. The velocity thus defined
can be measured in reasonable small systems (containing 200 and
600 particles respectively), while on the other hand being quite
close (within 10\%) from the large scale velocity. Based on this
definition of time of flight, we have a sound speed,
$c_\text{tof}=0.26$ in our units, at pressure $p=10^{-4}$.

In Fig.~\ref{fig:amplitude-width} we plot the scaling of the
amplitude and the width of the coherent part of the signal.  The
amplitude is well approximated with a power law, $A\sim
L^{-\gamma}$;  for the 2D simulations $\gamma \approx 1.5$. The
width of the coherent part of the signal increases with distance
also as a power law: $\sim L^\alpha$.  For the 2D simulations the
increase is close to linear, $\alpha\approx 1$.

We are not aware of any prediction or previous analysis of these
exponents $\gamma $ and $\alpha$ for polydisperse random packings.
In order to put these results into perspective, it is important to keep in
mind that $F_\text{sig}$ is not the amplitude of the wave motion in the
medium, but the resulting force on the boundary at the other edge.  Since
the force is proportional to the local stretching, i.e., the derivative of the
amplitude of the wave, $\gamma$ is not the exponent with which the wave
amplitude itself decays [see also the discussion at Eq.~(\ref{eq:wallforce})].

We have compared this behavior with the behavior of propagating pulses in
a one-dimensional chain of balls. Even in this simple system,
dispersion effects (wavenumber dependence of the frequency of the
waves) give rise to nontrivial exponents---as we shall discuss
in more detail in Sec.~\ref{sec:analytic}, both the exponent and
the shape of the pulse can be determined analytically. We collect
the exponents $\gamma $ and $\alpha$ in Table~\ref{tab}. An
important lesson from the 1D analysis is that the decay exponent
$\gamma$ is not universal, as it depends on the precise shape of
the initial pulse: $\gamma =2/3$ for our usual initial condition
(equilibrium position but nonzero velocity next to the wall at
$t=0$), and $\gamma =1$ if the initial condition is zero velocity
but nonzero displacement at the wall (not plotted).  If we allow
polydispersity in the 1D chain, the scaling appears to have a
larger exponent depending on the magnitude of the polydispersity,
although from the data shown in Fig.~\ref{fig:amplitude-width} we
cannot draw a definite conclusion.

In conclusion, the main qualitative differences between the 1D
results and those for the coherent pulse in the disordered 2D
packings is that {\em (i)} in the 1D chain the first pulse
broadens as $t^{1/3}$ whereas the pulse in the disordered 2D
medium broadens linearly; {\em (ii)} the amplitude of the pulse
decays much faster in the disordered medium than in the 1D chains
(in other words, $\gamma$ is larger).

\begin{table}
\caption{\label{tab}
  The scaling exponents $\gamma $ and $\alpha$ for different granular systems.
}
\begin{ruledtabular}
\begin{tabular}{lrr}
granular system  & $\gamma $ &$\alpha$\\
\hline
1D chain or triang.\ latt., monodisperse\footnotemark[1] & 2/3 & 1/3 \\
1D chain or triang.\ latt., monodisperse\footnotemark[2] & 1 & 1/3 \\
1D chain, polydisperse\footnotemark[1] (numerical) & $\ge 2/3$ & $\ge1/3$ \\
2D disordered\footnotemark[1] (numerical) & $\approx 1.5$ & $\approx 1$\\
\end{tabular}
\end{ruledtabular}
\footnotetext[1]{Initial condition (A): equilibrium position but nonzero
  velocity next to the wall at $t=0$}
\footnotetext[2]{Initial condition (B): zero velocity but nonzero displacement
  next to the wall at $t=0$}
\end{table}

\begin{figure*}
\resizebox{1.9\columnwidth}{!}{\includegraphics*{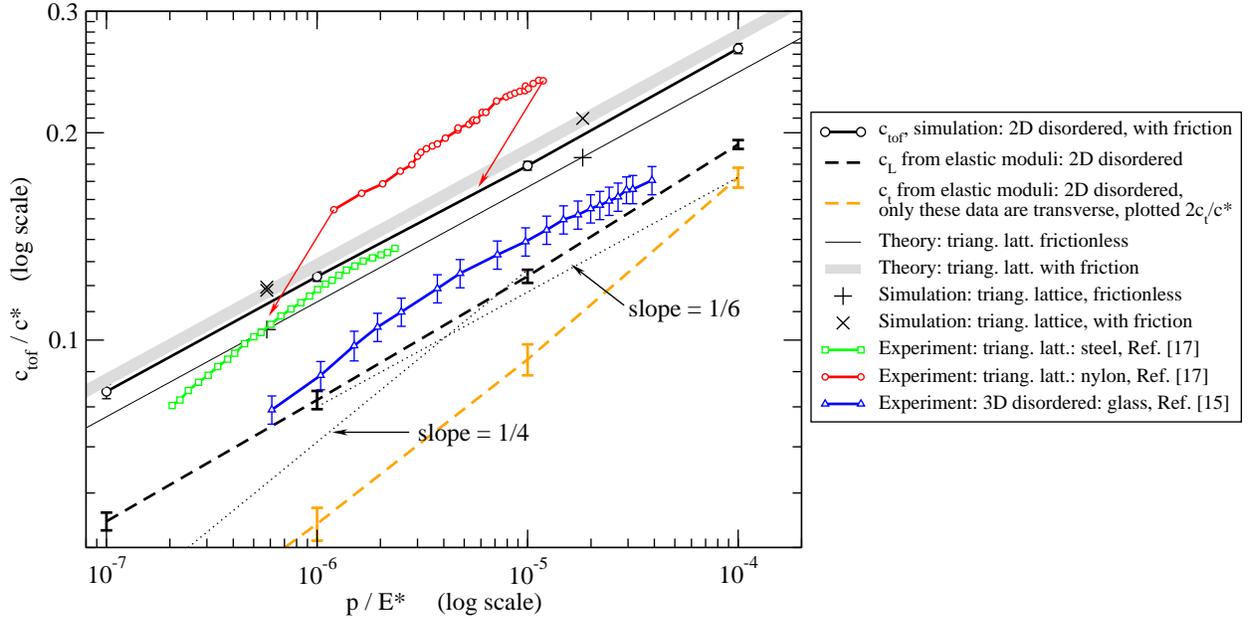}}
\caption{\label{fig:arrival-pressure}
  (Color online) The pressure dependence of the various sound
  speeds. The main data are the results for $c_\text{tof}$ obtained
  from our simulations, which show perfect $p^{1/6}$ scaling.
  Velocities $c_\ell$ and $c_\text{t}$ deduced from elastic moduli,
  as in~\eqref{eqn:vel_modu}, are smaller; one would have expected
  that $c_\text{tof} \sim c_\ell$.  $c_\text{t}$ is much smaller and has been multiplied by
  2 to fit within the scale of the plot.
 The theoretical curves for triangular lattice are
Eqs.~(\ref{eq:c-nofric}) (frictionless) and (\ref{eq:c-fric})
(frictional). For the latter there is a slight variation depending
on the Poisson ratio of the grain's material: the gray band
corresponds to the range $0\le\nu\le 0.5$, Simulations for the
frictionless triangular lattice ($+$) show excellent agreement
with Eq.~(\ref{eq:c-nofric}). For the frictional case ($\times$)
the simulation shows significant finite size scaling: it should
approach the top side of the gray band (we used $\nu=0$ and size
$L=24$, but for $p=5\times10^{-7}$ a larger system $L=160$ is also
plotted).  The simulation for 2D disordered frictional case
($\bigcirc$) shows results very similar to the triangular lattice.
For comparison we also show three experimental datasets [recall
the pressure- and velocity scales: $E^*=E/(1-\nu^2)$ and
$c^*=\sqrt{E^*/\rho}$ ]: triangular lattice of steel spheres
(green $\square$, from \cite{gilles03}), triangular lattice of
nylon balls (red $\bigcirc$, also from \cite{gilles03}; see text
for explanation of the arrows), and disordered 3D glass spheres
(blue $\triangle$ from \cite{jia99}).  For reference, lines with
slope $1/6$ (for the $p^{1/6}$ law) and $1/4$ (sometimes quoted as
effective exponent for low pressures) are shown. }
\end{figure*}

%\newpage
\subsection{Speed of sound, elastic moduli, and pressure dependence
\label{subsec:pressure}}

In this section we  turn our attention to the sound speed, and in
particular study its variation as a function of the confining
pressure $p$. The main quantity is the time-of-flight velocity
obtained from the propagation of the coherent pulse $c_\text{tof}$
(see section~\ref{subseq_coh}). 
It should be compared to the values of transversal ($c_\text{t}$) and
longitudinal ($c_\ell$) wave speeds that are deduced
(Eqn.~\ref{eqn:vel_modu}) from the apparent elastic moduli of
table~\ref{tab:moduli}.  We
also compare our results to experimental data for sound
propagation; since some experiments have been performed on regular
packings, we also have studied these analytically and numerically
(see section \ref{sec:analytic})
An overview of these various propagation velocities as function of
pressure is shown in Fig.~\ref{fig:arrival-pressure}. 
Let us first
discuss the scaling of $c_\text{tof}$, $c_\ell$ and $c_\text{t}$ (as
defined in~\eqref{eqn:vel_modu}) with $p$.
Recall that for a fixed contact network with Hertzian forces which
stay proportional to the pressure $p$, the sound velocity 
scales as $p^{1/6}$. We find here that
$c_\text{tof}$ follows this scaling quite accurately, while $c_\ell$
appears to be growing slightly faster as $p^{0.18}$. Surprisingly,
data for the velocity $c_\text{t}$ of transverse waves abide by a
different scaling, $c_\text{t} \sim p^{0.23}$; we do not know the reason
for this behavior.

Since the coherent wave is essentially longitudinal in nature, one
should compare $c_\ell$ and $c_\text{tof}$. Even though both
quantities scale rather similarly,  $c_\text{tof}$ is roughly 40\%
larger than $c_\ell$. As discussed in section~\ref{subseq_coh}, our
definition of $c_\text{tof}$ is based on measurements in
relatively small systems, and from a few simulations in larger
systems we found that this may overestimate $c_\text{tof}$ by some
10\%. In addition, if we do not measure the first arrival of the
signal, but instead measure the first peak location, or the first
zero crossing, $c_\text{tof}$ would go down substantially.

Furthermore, it seems that the pulse propagation with our method
of excitation does not probe the material on the longest scale. On
shorter scales, the material appears somewhat stiffer: As
discussed in appendix~\ref{app:elastic_local}, numerical
measurements of the elastic modulus $C_{22}$ can be performed on
various length scales, the shorter ones, in the case when displacements
are locally controlled, leading to larger apparent values of
$C_{22}$. 
In addition, we shall show in section \ref{subsec:modes} that there is
a strong contribution from non-plane-wave modes which cannot be
expected to be described by continuum elasticity.

It might therefore be concluded that a simple long-wavelength
description gives a good first approximation of the propagation
velocity of the coherent wavefront, but that modes that are not
accurately described by a long-wavelength approximation contribute
substantially to the wave propagation for 
the system sizes and excitation method employed here.
For a triangular lattice of monodisperse balls, we compare the
analytical expressions for the sound speed Eq.~(\ref{eq:c-nofric})
and (\ref{eq:c-fric}), which are derived in
section~\ref{sec:analytic} for infinitely large lattices, with
simulations on finite lattices. Both the frictionless and
frictional cases are in excellent agreement, even though  the
frictional one shows appreciable finite size corrections. The
simulation for 2D disordered frictional case shows results very
similar to the triangular lattice---including the $p^{1/6}$
scaling expected naively from the Hertzian force law---for the
range of pressures considered.

This quantitative agreement is somewhat surprising in view of the
large difference in coordination numbers (6 versus barely larger
than 3). This might partly be due to the small wavelength effects,
which affect the results in disordered systems, while one easily
observes the long wavelength result $c_\text{tof}=c_\ell$ with a
perfect regular lattice.

The only 2D data we are aware of are for spheres on a triangular lattice.
These systems are inevitably slightly polydisperse, which prevents the closing 
of all contacts between nearest neighbors on the lattice
\cite{velicky02,gilles03,roux97}, and in
the limit of low pressure, the coordination number should not exceed 4.
However, once the reduced pressure is high enough for the elastic deflection 
of contacts to compensate for the open gaps, the behavior of the perfect 
lattice is retrieved. This effect can be evaluated with the reduced pressure 
defined in \cite{roux97} as
\[
P^* =  {3P\over \alpha^{3/2} E^*}
\label{eqn:pstar}
\]
where $\alpha d$ is the width of the diameter distribution. Effects of 
polydispersity disappear as $P^*$ grows beyond 1. For steel spheres the data 
of \cite{gilles03} (for which $\alpha \sim 10^{-4}$) fall close to our
calculated values for the triangular lattice with friction when
$p\ge 10^{-6}$, as expected.
Even though there is a discrepancy in the velocity of the order of 10-20\%,
this agreement is remarkable, since $c_\text{tof}$ has been calculated without
any adjustable parameters. 
Possible finite size effects might explain why these
data lie below the theoretical frictional curve for the perfect
lattice.

The triangular lattice of nylon balls \cite{gilles03} shows
significantly larger rescaled velocity than expected. Possibly,
this discrepancy is simply a reflection of the uncertainty in the
effective elastic constant at the frequency range of the
experiments: nylon is a viscoelastic material for which the Young
modulus increases strongly with frequency. We do not know the
values of the elastic constants at the experiment's frequencies,
but nevertheless if we use a Young modulus twice as large as its
zero frequency value (for the plot the zero frequency modulus was
used), then the curve would shift as indicated by the arrows.

Finally, disordered 3D glass spheres \cite{jia99} display smaller
velocities than any 2D case. 
One possible explanation is that 2D experiments on planar sphere
assemblies can be viewed, if we imagine stacking such layers on top of
one another, as probing the stiffness or wave propagation along dense,
well coordinated planes in a 3D material with extreme anisotropy. This
renders plausible the observation of unusually high sound velocities,
in comparison with ordinary 3D packings.

\subsection{Eigenfrequencies and eigenmodes}\label{subsec:modes}

\begin{figure}
\resizebox{0.9\columnwidth}{!}{\includegraphics*{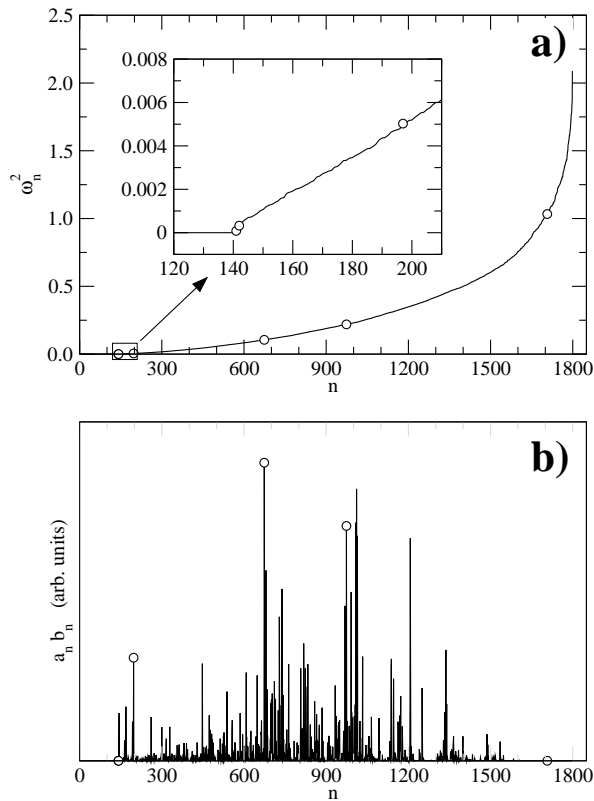}}
\caption{\label{fig:eigenvalues}
  \textbf{(a)} Eigenfrequencies of the linear system for the packing shown on
Fig.~\ref{fig:static}.
The squared eigenfrequencies are plotted against the number of the
mode $n$.  Modes $n=0,\ldots140$ have eigenvalue zero, as a
consequence of ``rattler'' grains which are not connected to the
network.  The \textbf{inset} shows a magnification of the plot
around the first few nonzero eigenvalues.
  \textbf{(b)} The contribution of the eigenmodes to the transmission
  signal $a_n b_n$ (see Eq.~\ref{eq:Ftop}).
On both panels the eigenmodes plotted on Fig.~\ref{fig:eigenmodes}
are marked by circles. }
\end{figure}

\begin{figure}
\resizebox{0.505\columnwidth}{!}{\includegraphics*{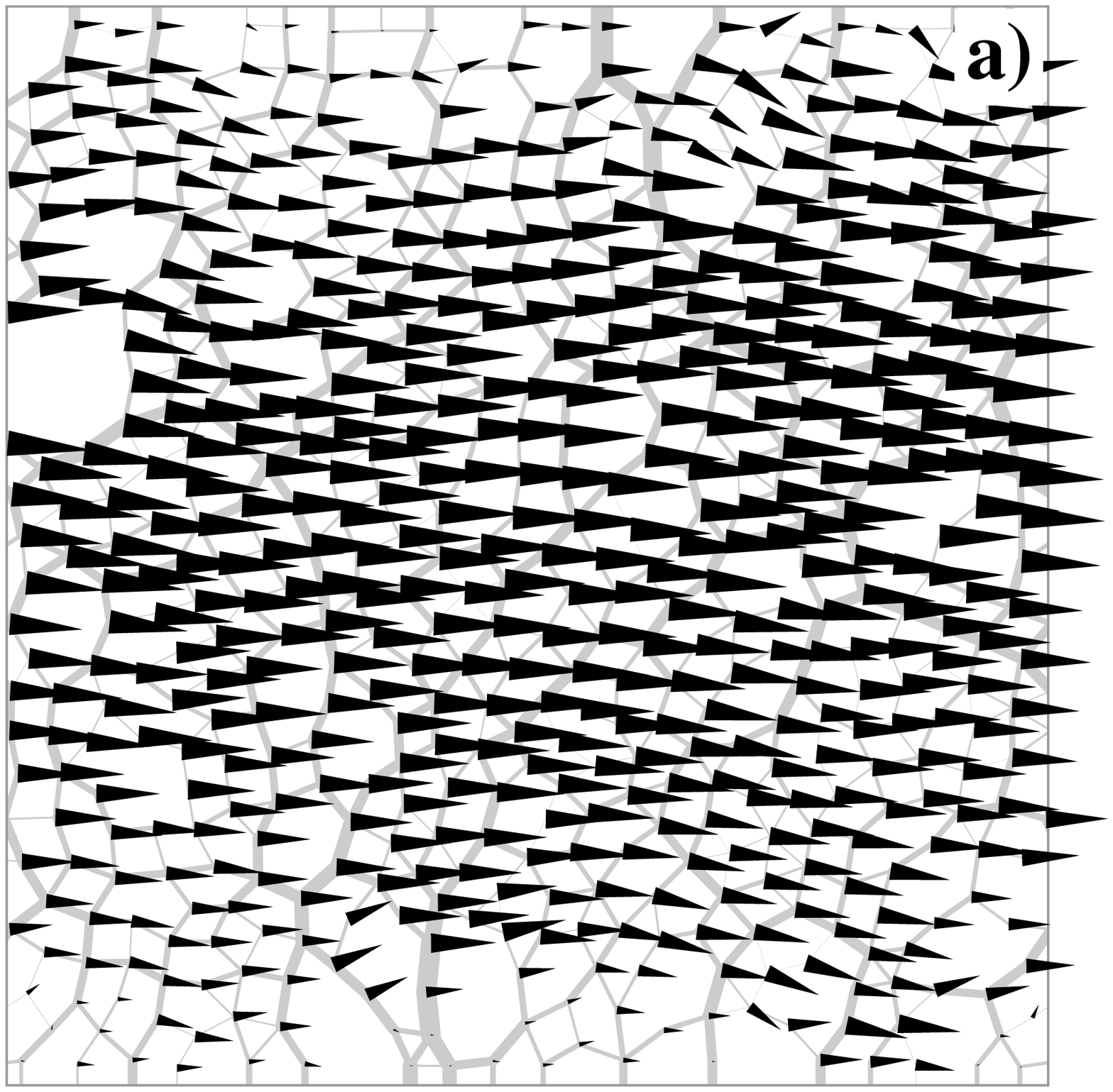}}
\hspace{-3mm}
\resizebox{0.505\columnwidth}{!}{\includegraphics*{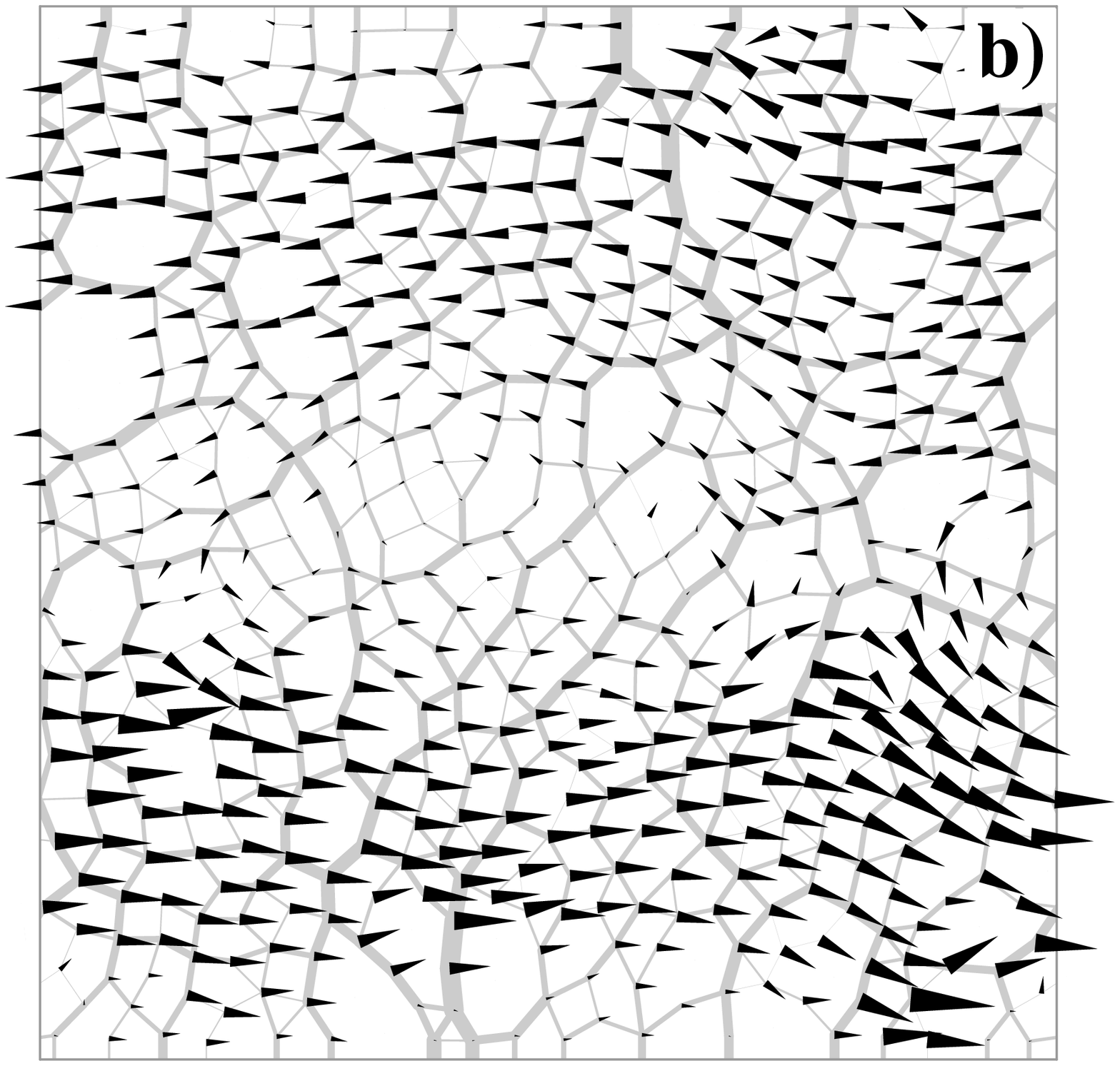}}
\vspace{-3mm}

\resizebox{0.505\columnwidth}{!}{\includegraphics*{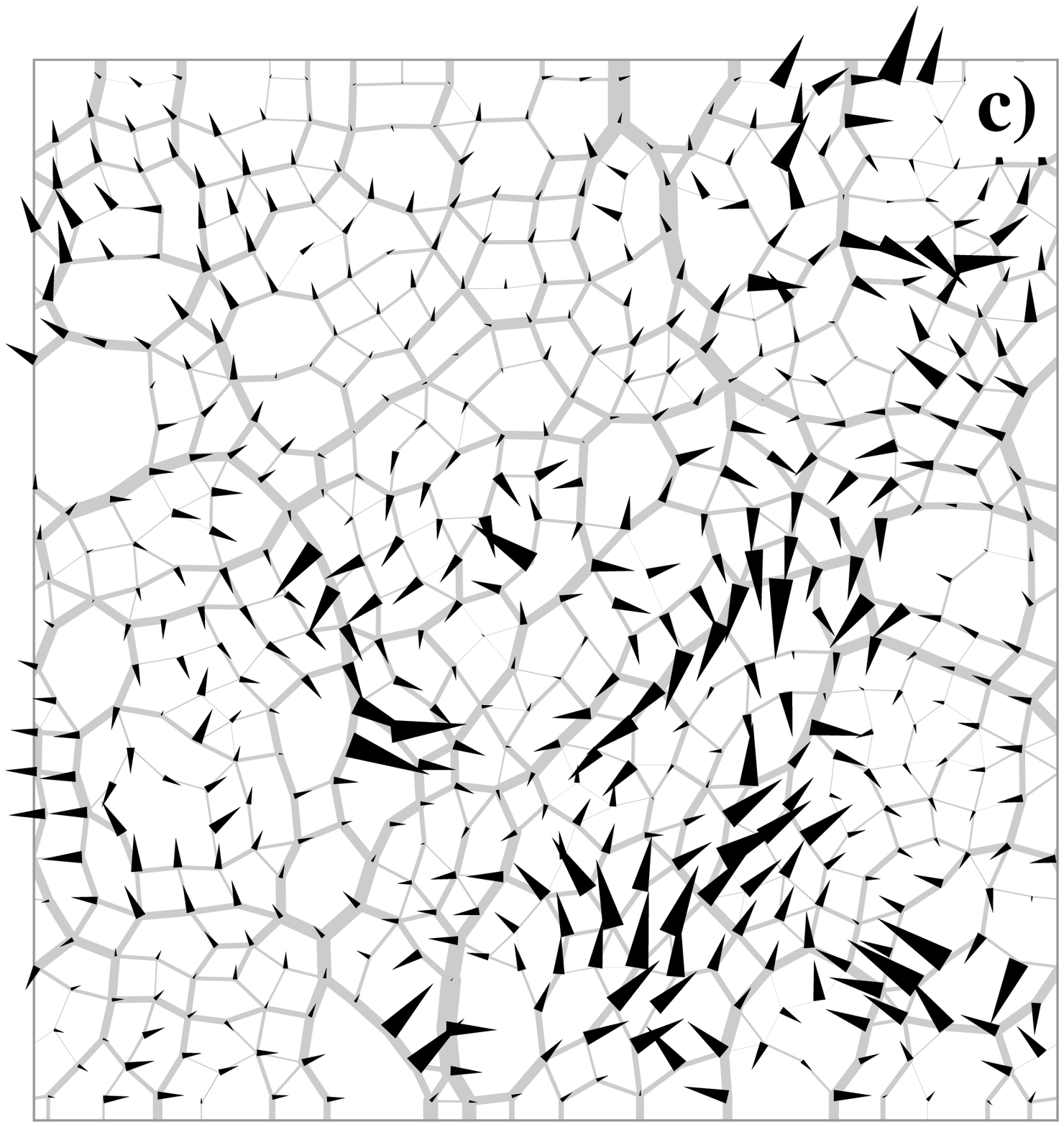}}
\hspace{-3mm}
\resizebox{0.505\columnwidth}{!}{\includegraphics*{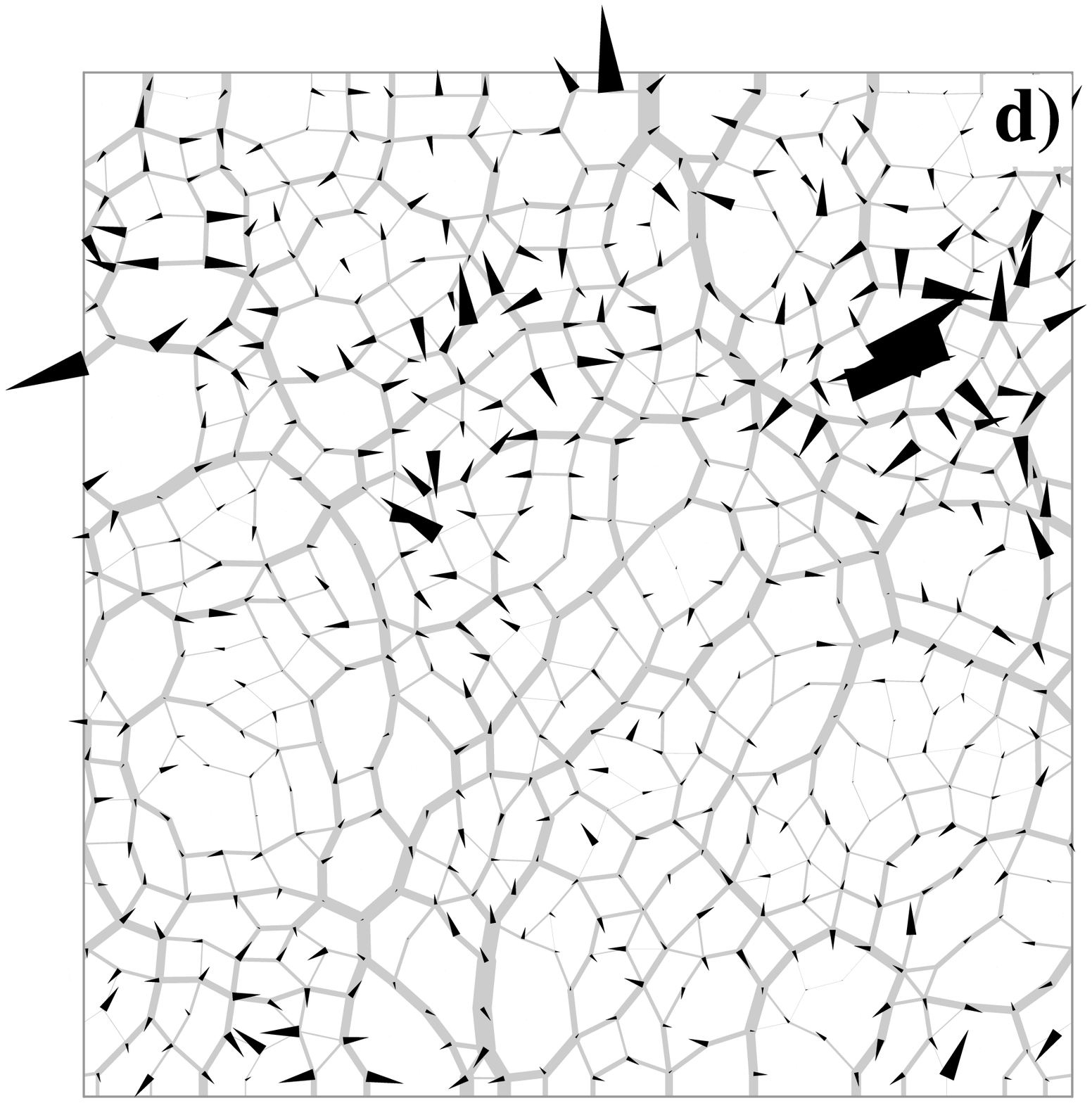}}
\vspace{-3mm}

\resizebox{0.505\columnwidth}{!}{\includegraphics*{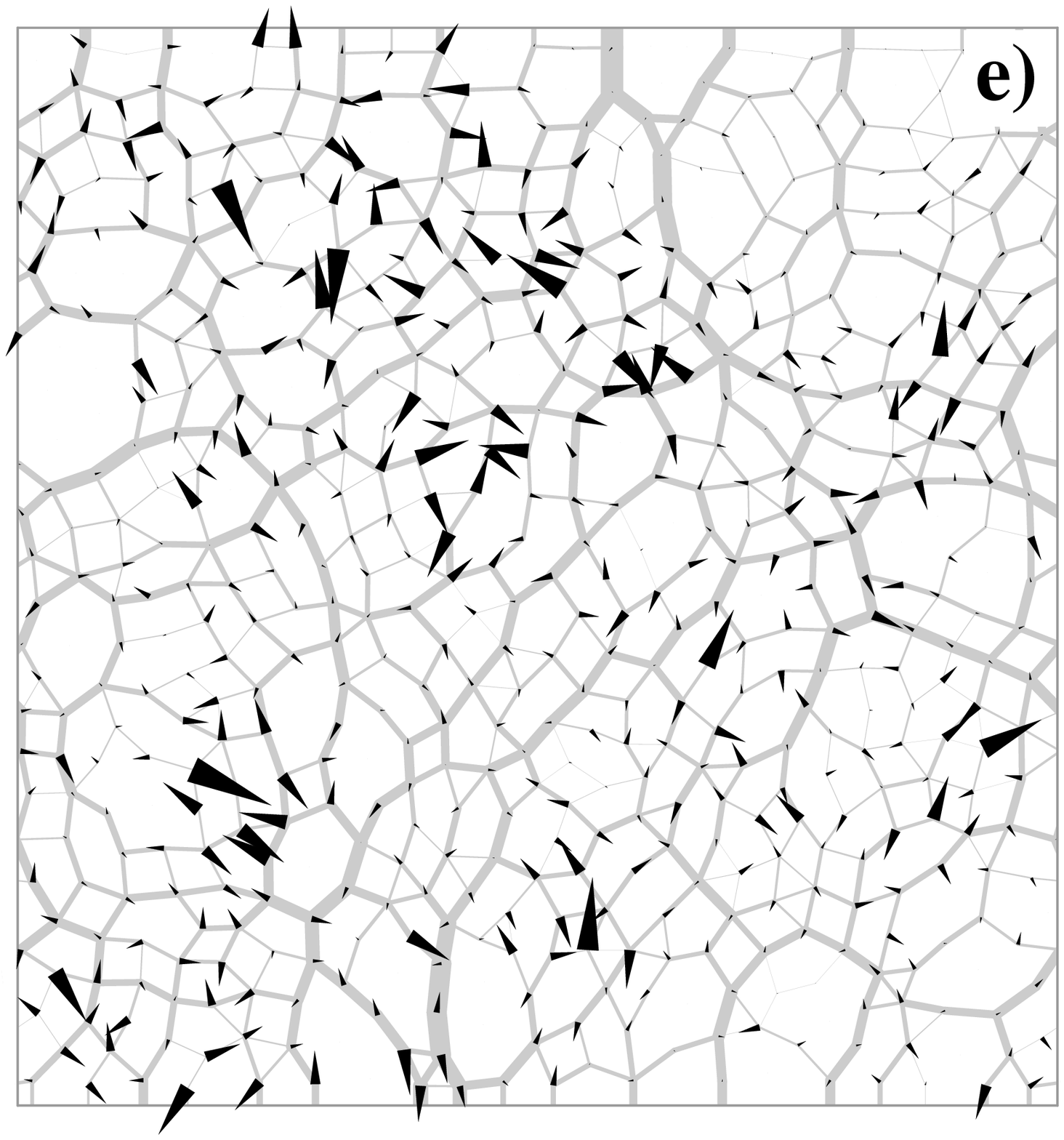}}
\hspace{-3mm}
\resizebox{0.505\columnwidth}{!}{\includegraphics*{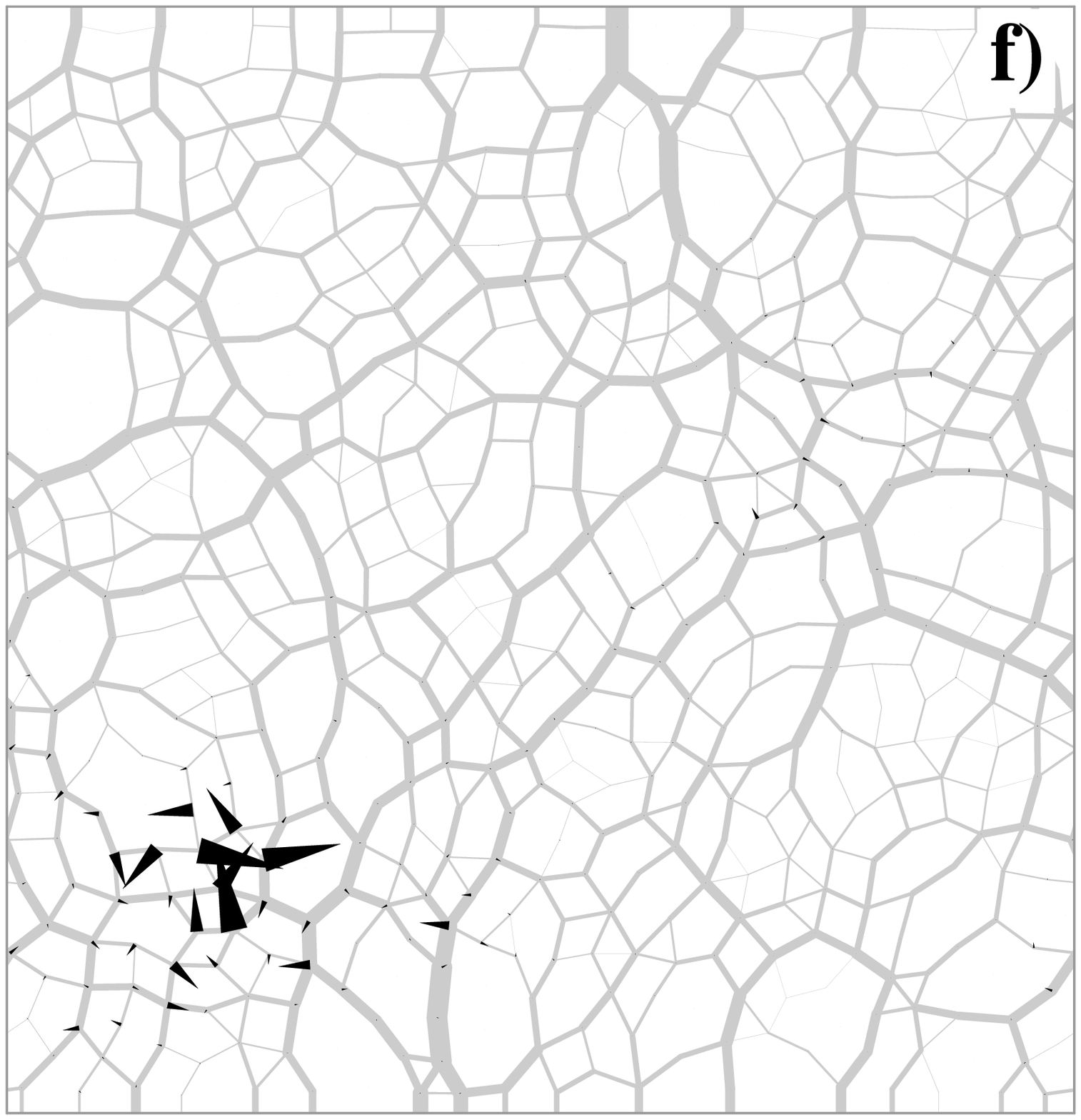}}
\vspace{-3mm}

\caption{\label{fig:eigenmodes}
  A few selected eigenmodes of the linear system. \textbf{(a)} $n=141$ and
\textbf{(b)} $n=142$ are the first two nonzero eigenmodes. They correspond to
the lowest excitation modes of a continuum body, though slightly distorted by
the disordered contact network.  \textbf{(c)} $n=197$, \textbf{(d)} $n=674$, and
\textbf{(e)} $n=974$ are some of the modes that contribute significantly to the
transmission of the signal.  \textbf{(f)} $n=1707$ is a high frequency
localized mode.  The modes shown here are marked on the eigenvalue plot,
Fig.~\ref{fig:eigenvalues}.
}
\end{figure}

In a rectangular sample of homogeneous elastic material with
boundary conditions similar to those employed here, the eigenmodes
are plane waves with wavevector $\mathbf{k}=(k_1,k_2)= (n_1
\frac{2\pi}{L_1},n_2 \frac{\pi}{L_2})$. If the tensor of elastic
moduli has the form given in~\eqref{eqn:moduli}, then it is
straightforward to show that the associated frequencies $\omega
_+$, $\omega_-$ are given by $\omega _\pm ^2 = \lambda _\pm
/\Phi_3^*$, $\lambda _\pm$ being the eigenvalues of the acoustic
tensor
$$
{\bf A}(k_1,k_2)=\bma C_{11} k_1^2+C_{33}k_2^2& (C_{12}+C_{33})k_1k_2 \\
(C_{12}+C_{33})k_1k_2 & C_{33}k_1^2+C_{22}k_2^2 \\ \ema,
$$
which implies that $\omega \propto k$ in the long wavelength
limit.

We show the spectrum of eigenmodes for a granular packing of 
600 grains at pressure $10^{-4}$ in Fig.~\ref{fig:eigenvalues}(a), and a few
selected eigenmodes in Fig.~\ref{fig:eigenmodes}.  There are a
number of zero eigenvalues because of ``rattler'' grains not
connected to the force network. The lowest nonzero modes
correspond to (slightly distorted) solid body modes, which are
similar to those expected from continuum theory. Remarkably, in
the absence of friction  it is much harder to identify eigenmodes
corresponding to continuum media modes, even for low frequencies;
and the low frequency modes are more abundant (not shown on the
figures). Nevertheless, the transmission signal looks rather
similar to the frictional case (Fig.~\ref{fig:signal}).

There are a large number of localized eigenmodes
[Fig.~\ref{fig:eigenmodes}(f)], which do not contribute
substantially to the signal transmission; clearly the modes that
dominate  the transmission  are global modes
[Fig.~\ref{fig:eigenvalues}(b)]: they contain oscillating grains
at both at the source and the detector wall. But with the
exception of only a few modes (with mode number 141-147 roughly),
their appearance is quite different from simple plane waves
[Fig.~\ref{fig:eigenmodes}(c-e)]. This indicates that, at least for
the system sizes, pressures and excitation method employed here,
the transmission of sound cannot be captured completely by
considering the material as a simple bulk elastic material. In
fact, in the light of these findings it is remarkable how close
the continuum prediction $c_\ell$ comes to $c_\text{tof}$. We will present
a more extensive study on these eigenmodes elsewhere
\cite{somfai040}.

\subsection{The effects of damping}
\label{subsec:damping}

\begin{figure}
\resizebox{0.9\columnwidth}{!}{\includegraphics*{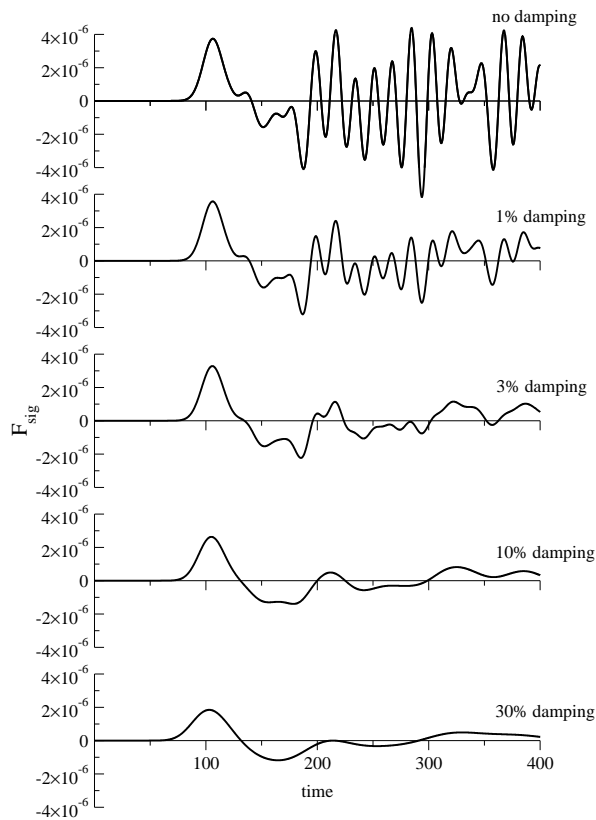}}
\caption{\label{fig:signal-damped}
  The transmitted signal with damping. The level of damping is expressed as a
fraction of the critical damping on each contact. The damping affects the
coherent part of the signal much less than the random part.
}
\end{figure}

Finally, we show here how damping affects the wave propagation. We
added viscous dissipation to the Hertz contacts in the way
described in appendix~\ref{Hertzapp}.  The resulting system cannot
be easily described by a linear model, and we obtained the wave
propagation signal by molecular dynamics simulations of the grain
oscillations. On Fig.~\ref{fig:signal-damped} we show the
transmission signal for a single configuration with various levels
of damping.  For large damping the coherent part of the signal is
only slightly altered, while the random part is strongly
suppressed.  This is in qualitative agreement with the experiments
of Ref.~\cite{jia01}, where damping was induced by adding a small
amount of water to the glassbead packing.

\section{Analytic results}
\label{sec:analytic}

\subsection{1D chain}

The problem of the propagation of a pulse in a 1D granular chain has been
considered by many authors \cite{nesterenko84,hinch99,chatterjee99,hascoet02,%
rosas03,rosas04,coste97} but the majority of the work is concentrated on
analyzing an initially uncompressed chain.  In this case the nonlinear force
law plays an important role, as well as the fact that there are no restoring
forces between the balls which initially just touch. For comparison with sound
propagation in granular media as a function of pressure, the relevant approach
is to first linearize the equations of motion starting from a compressed
chain, and then study the propagation of the pulse as governed by these
linearized equations.

The simplest system resembling the problem of sound propagation in granular
media (under pressure) is a 1D chain of identical
elastic balls, confined and compressed between two walls.  At $t=0$ we disturb
the first ball (see below for details), this disturbance travels with sound
speed $c$ in the chain, and arrives at the other wall at time $t_0=N\ell/c$,
where $\ell$ is the diameter of the balls.  For this system we can calculate
the scaling exponents and the waveform analytically.

In the Appendix we calculate the attenuation exponent of this wave.  For
initial condition (A), where the first ball has nonzero velocity but zero
displacement at $t=0$, the force with which the last ball presses the wall at
time $t_0$ scales with $N$ as
\begin{equation}
\label{eq:fA}
F^\text{A}(t_0) \sim N^{-2/3}\,.
\end{equation}
Initial condition (B), where all balls start with zero velocity but the first
has a finite displacement, gives a different answer:
\begin{equation}
\label{eq:fB}
F^\text{B}(t_0) \sim N^{-1}\,.
\end{equation}
These are the attenuation exponents for a uniform 1D chain.

To derive the waveform analytically in the large system and long time limit,
we consider the long wavelength expansion of the dispersion relation
(\ref{eq:dispersion}):
\begin{equation}
\label{eq:disp3}
\omega_n \approx ck_n - \frac{c\ell^2}{24}k_n^3\,.
\end{equation}
A propagating wave solution $u(x,t)=A \exp(ikx-\omega t)$, where for long
wavelengths $x$ can be considered continuous, has to satisfy the following
partial differential equation to match dispersion relation (\ref{eq:disp3}):
\begin{equation}
-\frac{\p u}{\p t} = c\frac{\p u}{\p x} + \frac{c\ell^2}{24} \frac{\p^3u}
{\p x^3}
\end{equation}
Changing variables to the co-moving frame, $\xi=x-ct$, the $\p u/\p x$ term
drops out.  Looking at similarity solutions of the form
\begin{equation}
u(\xi,t)\sim t^{-g} U\left(w=\frac{\xi}{t^\alpha}\right)\,,
\end{equation}
we obtain $\alpha=1/3$ and
\begin{equation}
\label{eq:simeq}
0=-g U(w) -\frac{w}{3}U'(w) + \frac{c\ell^2}{24}U'''(w)\,.
\end{equation}
This leads to different classes of solutions for different attenuation
exponent $g $.  First we consider the case $g =0$, which leads to Airy
functions: $U_0'(w) = \Ai({2w}/{(c\ell^2)^{1/3}})$.  We can generate further solutions
by differentiating Eq.~(\ref{eq:simeq}).  For example by differentiating once
and twice we find that $U=U_0'$ solves Eq.~(\ref{eq:simeq}) for
$g =1/3$, while $U_0''$ solves it for $g =2/3$. 
For $u(x,t)$ this gives solutions, e.g., for $g=1/3$ as
$t^{-1/3} \Ai[2(x-ct)/(c\ell^2 t)^{1/3}]$, which as we show soon is the
selected solution for initial condition (A). At this point we need more
information to see which solution is selected: Eqs.~(\ref{eq:fA}) and
(\ref{eq:fB}) tell the exponent of the scaling of the force at the wall with
$N$.  Note that $u(x,t)$, is the propagating solution in a semi-infinite
medium; the solution for a reflecting wall boundary condition,
$u\big(x=(N+1)\ell,t\big)=0$, is composed of two counter-propagating waves.
Now the force at the wall is proportional to the displacement
of the ball (at $x=N\ell$) next to the reflecting wall [at $x=(N+1)\ell$],
hence it can be written as
\begin{equation}
\label{eq:wallforce}
F(t) = K\big\{u(x,t)-u(2(N+1)\ell-x,t)\big\} \,,
\end{equation}
which with $x=N\ell$ and $t=t_0+\tau$ gives for the $g=1/3$ solution
\begin{eqnarray}
\label{eq:airy}
\nonumber
F%^\text{A}
(t_0+\tau)
&\sim& K\ell \left(\frac{ct}{\ell}\right)^{-1/3} \left\{
  \Ai\left(\frac{-2c\tau}{(c\ell^2t)^{1/3}}\right)- \right.
\\
&& \qquad \left. - \Ai\left(\frac{4\ell-2c\tau}{(c\ell^2t)^{1/3}}\right)\right\}
\\
&\sim& -4K\ell \left(N+\frac{c\tau}{\ell}\right)^{-2/3}
  \Ai'\left(\frac{-2c\tau/\ell}{(N+c\tau/\ell)^{1/3}}\right).
\nonumber
\end{eqnarray}
Note that because of the extra differentiation the decay exponent of the force
on $N$ does not equal $g=1/3$ but instead becomes $\gamma=g+1/3=2/3$.
This scaling exponent is the same as for the initial condition (A) in
Eq.~(\ref{eq:fA}), showing that indeed the $g=1/3$ solution is selected
here.  

To match the initial condition (B), we need to use the $g=2/3$ solution:
\begin{equation}
\nonumber
F^\text{B}(t_0+\tau)
\sim -4K\ell \left(N+\frac{c\tau}{\ell}\right)^{-1}
  \Ai''\left(\frac{-2c\tau/\ell}{(N+c\tau/\ell)^{1/3}}\right).
\end{equation}

\begin{figure}
\resizebox{0.99\columnwidth}{!}{\includegraphics*{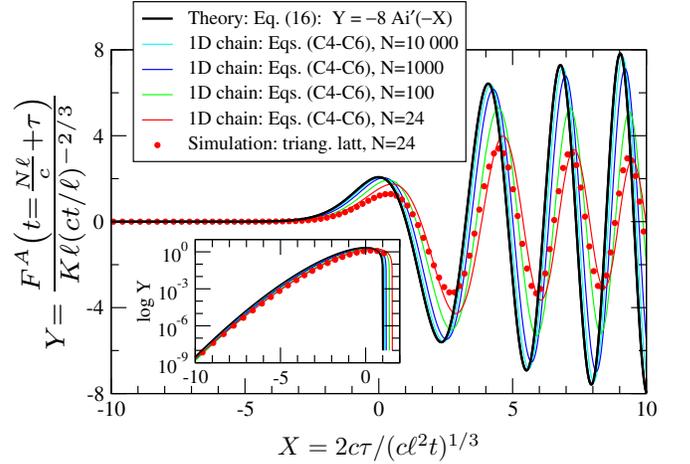}}
\caption{\label{fig:1D}
  (Color online) Comparison of signal shapes for initial condition (A). The
theoretical prediction of Eq.~(\ref{eq:airy}) (thick black line) is compared
to the 1D chain of identical balls, numerical sum in
Eqs.~(\ref{eq:soln}--\ref{eq:force}) (color lines).  The simulation
of a perfect triangular lattice (red full circles) is very close to the 1D
chain of the same size.
Note that Eq.~(\ref{eq:airy}) has an undetermined multiplicative factor as it
is obtained as a solution of a linear equation.
}
\end{figure}

The solution of the 1D chain, obtained by numerically evaluating the sum
(\ref{eq:soln}) converges for $N\to\infty$ to the analytical solution, see
Fig.~\ref{fig:1D} where the initial condition (A) is plotted.

At this point we can understand the connection between the two initial
conditions.  Initial condition (B) is related to (A) by a time differentiation.
Since the equations are linear, the solutions are similarly related to each
other.  The above solutions have the  structure that differentiating one of them
and dropping  subdominant terms gives another of the solutions, with exponent
$g $ which has increased  by $1/3$.

If we allow for disorder in the 1D chain, the results change slightly.
We introduce disorder by varying the radii of the ball, and solve this 1D
system numerically, as described in Sec.~\ref{sec:model} for 2D and 3D
packings.  On Fig.~\ref{fig:amplitude-width} we show the exponents of the
polydisperse 1D chain.  Both the amplitude exponent $\gamma $ and the width
exponent
$\alpha$ appear to be larger than in the case of identical balls, but the
results are not clear enough to extract a value for the exponents.

The above analysis shows that in the long time limit, the propagation of an
initially localized pulse is governed by an Airy equation---as
Fig.~\ref{fig:1D} shows, the first pulse and the first oscillations behind it
converge to an Airy function type behavior when viewed in a frame co-moving
with the initial pulse. Note that the kinetic energy in the leading pulse
decays rapidly as $t^{-2g-1/3}$:
\begin{eqnarray}
E_\text{kin}&=&\sum_\text{init.pulse}\frac{m}{2} \dot u_n^2(t)
\,\simeq\, \frac{m}{2}\sum_\text{init.pulse}c^2\frac{t^{-2g}}{t^{2/3}}\left[
U'(w_n)\right]^2\nonumber\\
&\sim& \frac{t^{-2g}}{t^{2/3}} t^{1/3} \sim t^{-2g-1/3} \,,
\end{eqnarray}
because the number of terms in the sum which contribute to the first peak is
proportional to the width of the pulse, which scales as $t^{1/3}$.  Hence for
the pulse shown in Fig.~\ref{fig:1D}, the kinetic energy in the first pulse
decays as $t^{-1}$, since $g=1/3$.  This illustrates that as time progresses,
more and more of the energy is stored in the region behind the first pulse.
The oscillations in this region are relatively incoherent, with a frequency
comparable to the maximum frequency of the dispersion relation. As the size of
this region increases linearly with time, the typical amplitude of these
oscillations decays as $t^{-1/2}$.  One can also obtain a $t^{-1/2}$ type
decay directly from a steepest descent analysis near the maximum of the
dispersion relation of the linearized equations of motion.

The fact that the first pulse in 1D chains is described by an Airy function
has been noted before \cite{hinch99,rosas03}.  Most of these studies are for 
initially uncompressed chains, however.  In this case, all the energy remains
confined in the first pulse, due to the absence of restoring forces.  As a
result, the exponent of the time dependence of the amplitude is different, and
consistent with energy conservation in the leading pulse.

\subsection{Triangular lattice}
In view of the experiments of Gilles and Coste \cite{gilles03}, it is
illuminating to also apply these results to  the simplest 2D system: a
triangular lattice of balls, with
rectangular boundaries.  The initial condition is given on balls touching one
wall of the rectangle, and we assume that a lattice direction of the
triangular lattice is parallel to this wall.
The longitudinal sound speed in a perfect triangular lattice (no
polydispersity) of Hertzian balls can be easily calculated, see, e.g.,
\cite{velicky02}. For the frictionless and frictional case respectively it is
given by
\begin{eqnarray}
\label{eq:c-nofric}
\frac{c^\text{nofric}}{\sqrt{E^*/\rho}} &=& \frac{3^{19/12}}{2^{3/2}\pi^{1/2}}
  \left(\frac{p}{E^*}\right)^{1/6} ,\\
\label{eq:c-fric}
\frac{c^\text{fric}}{\sqrt{E^*/\rho}} &=& \frac{3^{19/12}}{2^{3/2}\pi^{1/2}}
 \sqrt{1+\frac{\eta}{3}} \left(\frac{p}{E^*}\right)^{1/6} \,,
\end{eqnarray}
where the parameter $\eta$ is the ratio of the tangential and radial
stiffnesses of a Hertz-Mindlin contact, see Eq.~(\ref{eq:eta}).

One way to calculate this is to map to the 1D chain of identical balls.  If in
the triangular lattice the longitudinal motion is perpendicular to rows, then
a row of $M$ balls moves together, corresponding to a single ball in 1D.  Thus
of the $N$ rows each has mass $m_\text{eff}=M\rho\pi/6$, they are separated by
distance $\ell=\sqrt{3}/2$ (recall our length unit was the ball diameter), and
connected by an effective spring $K_\text{eff}=3^{7/6} 2^{-2} M p^{1/3}$.  In
the frictional case $K_\text{eff}$ has an additional factor of $(1+\eta/3)$.

This way we also predict the shape of the signal for the triangular lattice,
see Fig.~\ref{fig:1D}.  The cause of the slight deviation from the 1D chain
result is a consequence of the fact that in the triangular lattice the springs
connecting to the walls are different:
$K_\text{wall}/K_\text{bulk}=3^{5/6}/4 \approx 0.62$.

If the radii of the balls are polydisperse, then at pressures low enough (that
the length scale of the elastic deformations become comparable to the
polydispersity) the stress field fluctuates spatially.  The effect of this on
the sound speed has been calculated by Velick\'y and Caroli \cite{velicky02}
in a  mean-field approximation.

\section{Discussion}
\label{sec:summary}

We presented numerical simulations of pulse propagation in 1D, 2D
and 3D granular systems. This response can be decomposed into an
initial coherent part, which is independent on the details of the
packing, and a subsequent random part, which is strongly
realization dependent. We have focused on the properties of the
initial coherent front. Our first observation is that the response
to a pulse propagates linearly in time, defining a time-of-flight
velocity, and does not follow force chains.

The fact that the packings in our numerical simulations have
roughly the same number of grains per container side (although in
2D) as the systems which have been studied experimentally by Jia
\emph{et al.} \cite{jia99}, and that our temporal signals are very
comparable to the experimental ones, makes us confident that our
simulation results can be fruitfully compared to experiments like
these. Indeed we find that the 3D experimental and 2D numerical
results for the time-of-flight velocity are in reasonably good
agreement. The experiments in 2D are done on triangular lattices,
and we also study numerically and analytically pulse propagation
on such lattices with and without friction. The experiments for
steel spheres and the predictions for frictional lattices are in
good agreement (even though there are some subtle points regarding
the scaling with pressure, see below).

We also compare our numerical results for the disordered system
with predictions following from numerical estimates of the
effective long wavelength elastic constants of our packings.
Remarkably, even though elastic constants predict the
time-of-flight velocity reasonably well, there is a 40 \%
discrepancy between predicted and observed velocity for our
systems. A possible reason for this is that our pulses may probe
the system on short scales which are not governed by a long
wavelength expansion; indeed a (preliminary) analysis of the
spatial structure of the modes that contribute significantly to
pulse propagation indicates that most modes appear rather
different from simple plane waves. It is likely, but
very hard to check numerically (at least with the methods used in this paper),
that for propagation over larger
distances (such as those probed in the engineering literature) the
elastic approximation becomes better, and the dominant modes would
become simple plane waves. The crucial open question becomes thus
what sets the length scale at which such description becomes
applicable. Recently this issue has also emerged within the
context of the proposal that the static behavior of granular
packings of hard particles is governed by a critical point
(``point J'') in a jamming phase diagram \cite{ohern03}. We will
come back to the relation to the jamming phase diagram elsewhere
\cite{somfai040}.

We also found that the amplitude of the coherent part and its
width scales with a power of the distance as the signal
propagates. For the initial condition where grains touching one
wall have nonzero velocity, the amplitude exponent is $\gamma
\approx 1.5$ for disordered 2D systems, while it is $2/3$ (exact
result) for a 1D chain of identical balls. The exponent of the
signal width is $\alpha\approx1$ for the disordered 2D system, and
$1/3$ for the 1D chain. The shape of the signal can be computed as
well, and it is given by Airy functions for the 1D chain.  A
triangular lattice of identical balls can be mapped (except for
the strength of the wall springs) to the 1D chain, predicting the
same exponents and signal shape.

A final issue that we studied in detail is the variation of the
sound velocity with pressure, since this is an important
experimental parameter. 
Our simulations for frictional contacts recover 
the expected $p^{1/6}$ behavior
for the time-of-flight velocity and
bulk modulus, but not for the shear modulus: we found that the transversal
wave speed scales approximately as
$p^{1/4}$. These results should be compared to results for
frictionless sphere packings with Hertzian contacts as studied by
O'Hern {\em et al.} \cite{ohern03}. They found that the bulk modulus $B\sim
p^{1/3}$ at
low pressure, while the shear modulus $G$ scaled as $p^{2/3}$,
resulting in ${\displaystyle c_\ell = \sqrt{\frac{B+(4/3)G}{\rho}}
\sim p^{1/6}}$ and $ {\displaystyle c_\text{t} = \sqrt{\frac{G}{\rho}}
\sim p^{1/3}}$.

Some of the experimental data for $c_\text{tof}$
~\cite{jia99,jia01,gilles03} or for resonance frequencies~\cite{DM57}
in bead assemblies, and some evaluations
of elastic moduli in numerical simulations~\cite{roux97,makse99},
evidence a larger exponent, or at least some departure from the
$p^{1/6}$ scaling. The physical origin of such observations has been
the subject of considerable debate~\cite{GO90,makse99,roux97,velicky02}. 
In fact, results for the pressure dependence of sound velocity in
disordered glass bead packings are somewhat different according to the
conditions of the experiment, and apparent values of exponents
$\alpha$ in a $c\sim p^\alpha$ fit vary roughly between $0.16$ and
$0.25$.
This calls for detailed investigations of the influence of the internal
state of packings on sound velocities and their pressure dependence.
While the data published in~\cite{jia99}, shown on
Fig.~\ref{fig:arrival-pressure}, indicate a crossover from
$p^{1/4}$ at low $p$ to $p^{1/6}$ at higher pressure, Domenico's
results~\cite{DO77}, corresponding to much larger confining
stress, are fitted by a $p^{1/4}$ law. Other data by Jia and Mills
(see \emph{e.g.,} ref.~\cite{jia01}, figure 10) agree with $c\sim
p^{0.21}$ on the whole studied pressure range, while Sharifipour
\emph{et al.}~\cite{SDH04} report in some cases exponents $\alpha$
as high as $0.28$. 
Pressure dependences with exponents $\alpha \sim 0.25$ often observed
with sands (see~\emph{e.g.,} \cite{HI96}) are likely to
 be related to the non-Hertzian behavior of contacts between~\emph{angular}
particles (or between asperities of rough particles) as discussed by
Goddard~\cite{GO90}, and are outside the scope of our simulations. 

Another suggested origin for a different effective scaling for
Hertzian contacts is the increase of coordination number with
pressure~\cite{GO90,roux97,makse99,ohern03,gilles03}, which gradually
stiffens the packing. In our case, this increase is rather small (from $z^*
\simeq 3.2$ to $z^* \simeq 3.5$) and does not entail any deviation from
the $p^{1/6}$  scaling for the effective longitudinal speed $c_\text{tof}$.

Such an explanation by pressure-induced recruitment of additional
contacts seems more plausible in regular lattices of nominally identical
spheres~\cite{DM57,roux97,velicky02,gilles03}, in which a slight
polydispersity (or lack of sphericity) causes lattice imperfections
and strongly reduces the coordination number,
which only recovers the perfect 
lattice value at high enough confining pressure, $P^*\ge 1$ (see 
\eqref{eqn:pstar} ).
We can expect such a
mechanism to explain the experimental observations by Duffy and
Mindlin~\cite{DM57} and Gilles and Coste~\cite{gilles03}, as
numerical studies of elastic moduli~\cite{roux97}, as well as a
self-consistent ``effective medium'' approach by Velick\'y and
Caroli~\cite{velicky02}, both find deviations from a $p^{1/6}$ scaling
of long wavelength sound. This effect is of course absent in our
simulations of perfect lattices.

\begin{acknowledgments}
This research has been supported by the Dutch FOM Foundation and the PHYNECS
training network of the European Commission under contract HPRN-CT-2002-00312.
MvH acknowledges support from NWO. \\
Laboratoire des Mat\'eriaux et des Structures du G\'enie Civil is a
joint laboratory, depending on Laboratoire Central des Ponts et
Chauss\'ees, Ecole Nationale des Ponts et Chauss\'ees and Centre
National de la Recherche Scientifique. \\
We thank Christophe Coste, Martin Depken, Chay Goldenberg, Ray Goldstein,
Xiaoping Jia, Bert Peletier, Adriana Pesci, and Tam\'as Unger for discussions.
\end{acknowledgments}

\appendix

\section{Contact forces in numerical model}\label{Hertzapp}

The normal force between two particles in contact is given by the
3D Hertz law \cite{johnson85}, which is the force between two
elastic spheres (labeled 1 and 2):
\begin{equation}
F_\text{n} = \frac{4}{3}\sqrt{R_{12}}E_{12}^* n^{3/2}\,,
\end{equation}
where the effective radius $R_{12}=[(R_1)^{-1}+(R_2)^{-1}]^{-1}$
and effective Young modulus $E_{12}^* =
[(E^*_1)^{-1}+(E^*_2)^{-1}]^{-1}$ are half of the harmonic
averages of the two grain's parameters.  Here we introduced the
material parameter $E^*=E/(1-\nu^2)$ (modified Young modulus,
non-standard notation), where $E$ is the Young modulus and $\nu$
is the Poisson number.  The distance of approach (or ``virtual
normal overlap'') is given by $n=R_1+R_2-r_{12}$, where $r_{12}$
is the distance of the two particle centers.  Grain--wall
interaction can be obtained setting $R_\text{wall}=\infty$, and we
used hard walls ($E_\text{wall}=\infty$).

Implementing the frictional force is less straightforward, because
frictional contacts can have a memory of their history. The
standard approach is to consider changes in the tangential force
with Mindlin's approximation \cite{johnson85}:
\begin{equation}
\label{eq:fricforce} \Delta F_\text{t} = 8
G_{12}^*\sqrt{R_{12}n}\Delta t \,,
\end{equation}
where the elastic constant
$G_{12}^*=[(G^*_1)^{-1}+(G^*_2)^{-1}]^{-1}$ can be calculated from
the two grains' material parameter $G^* = E/[2(1+\nu)(2-\nu)]
=E^*(1-\nu)/(4-2\nu)$, and the virtual tangential displacement $t$
of the particle surfaces is determined from their centers' motion
and their rotations.  This incremental force law is augmented with
the Coulomb condition:
\begin{equation}
|F_\text{t}| \le \mu F_\text{n} \,,
\end{equation}
where we take the friction coefficient $\mu$ as parameter.  It
is interesting to note that for a given contact the ratio of the
normal stiffness to the tangential stiffness (despite the very
different force laws) is constant, we call it $\eta$:
\begin{equation}
\label{eq:eta} \eta := \frac{d F_\text{t} / dt}{d F_\text{n} / dn}
= \frac{8 G^*}{2 E^*}= 1-\frac{\nu}{2-\nu} \,,
\end{equation}
assuming the two grains have the same elastic parameters. For
example for $\nu=0$, the value we use in most of the simulations,
the two effective stiffnesses $ d F_\text{t} / dt $ and $d
F_\text{n} / dn$ are equal, so for vibrations neither the radial
nor the frictional part of the contact will dominate the other.
(Note that Eq.~(\ref{eq:eta}), a consequence of approximation
(\ref{eq:fricforce}), is really only valid for $|F_\text{t}| \ll
\mu F_\text{n}$, as pointed out in \cite{velicky02}.)

The Coulomb condition introduces dissipation, because contact
surfaces may slip at nonzero force. The ensuing dissipation occurs
only when the yielding threshold is exceeded, and not in the
infinitesimal amplitude oscillations we study here. Nevertheless,
in some cases we wish to add dissipation, for example when
creating the packing from the granular gas, or when studying the
effect of damping on the small amplitude oscillations.  For this
purpose we chose a particular form of damping, which is {\em at
every instant} a given constant fraction of the linear critical
viscous damping, both for the normal and for the tangential force.
Through this procedure, the effective  damping force, like the
normal and tangential Hertz-Mindlin forces, depends nonlinearly on
the distance of approach $n$.  We impose that the total radial force, which
now also contains the viscous contribution, never becomes attractive.

This choice is appealing theoretically because it contains only
one non-dimensional parameter to control the strength of the
dissipation.  In practice it is not clear what the best
approximation is of the real (dry or wet grains') dissipation.  In
any case there should be some contacts that dominate the
dissipation.  For those contacts the viscous force is a certain
fraction of the critical damping. Our approach is that we impose
this ratio on \emph{all} contacts.

\section{Small scale static elasticity \label{app:elastic_local}}

Let us consider a homogeneous macroscopic sample of an elastic
material with the same symmetries and boundary conditions as our numerical systems, and apply
a body force (per unit volume):
\be
f_\alpha ^{(n)} (y) = f_0 \sin (\frac{n \pi y }{L_2}),
\label{eqn:fnalpha}
\ee
depending on coordinate $y$, and directed parallel to axis $\alpha$ (1
or 2). With boundary conditions ${\bf u}=0$ on the top ($y=L_2$) and
bottom ($y=0$) walls, and lateral periodicity ($L_1$ is the system
width) the corresponding displacement field only has a
non-vanishing coordinate $u_\alpha$, given by:
$$
u_\alpha ^{(n)}(y) = \frac{f_0 L_2^2}{n^2 \pi^2 C_{\alpha}}
\sin(\frac{n \pi y }{L_2}),$$
with $C_{\alpha} = C_{22}$ for $\alpha=2$ and $C_{\alpha} = C_{33}$
for $\alpha=1$. Hence the total elastic energy
\be
\tilde W_\alpha ^{(n)} = \frac{L_1 L_2^3 f_0^2}{4 n^2\pi^2C_\alpha}.
\label{eqn:wnalpha}
\ee
To mimic the force field of Eqn.~\ref{eqn:fnalpha} in our discrete
samples, each non-rattler bead $i$ is submitted to a force~:
$$F^{(n)} _{i,\alpha} = \frac{4\pi R_i ^3} {3\Phi^* _3} f_0 \sin
(\frac{n \pi y_i }{L_2}),$$
while the bottom and top walls are fixed and an apparent elastic
modulus is obtained from the total elastic energy using
Eqn.~\ref{eqn:wnalpha}. The resulting ``local'' constants, denoted as $\tilde
C_{22}(n)$ and $\tilde C_{33}(n)$ are compared in
table~\ref{tab:modul2}, for $n=1$ and $n=2$, to the corresponding ``global'' values
(given in table~\ref{tab:moduli}). The longitudinal constant $C_{22}$,
  measured in this way, is only slightly lower, but roughly agrees with
  the previous result. Results for transverse constants are similar,
  except that sample to sample fluctuations are somewhat larger. In view of the small sample size, and the
  importance of boundary effects, it can be concluded, especially for
  the higher pressures, that the static elastic response to force fields
is in reasonable agreement with the equations of elasticity involving the moduli
measured on globally deforming the sample.
\begin{table}
\centering
\begin{tabular}{|c|cccc|}  %\cline{1-5}
\hline
$p$&$\tilde C_{22}(1)/C_{22}$&$\tilde C_{22}(2)/C_{22}$&$\tilde C_{33}(1)/C_{33}$&$\tilde C_{33}(2)/C_{33}$\\
\hline
$10^{-7}$
&$0.85\pm 0.05$
&$0.74\pm 0.09$
&$0.81\pm 0.14$
&$1.01\pm 0.24$\\
\hline
$10^{-6}$
&$0.88\pm 0.04$
&$0.80\pm 0.06$
&$0.90\pm 0.20$
&$0.96 \pm 0.16$\\
\hline
$10^{-5}$
&$0.91\pm 0.05$
&$0.86\pm 0.06$
&$0.85\pm 0.10$
&$0.89\pm 0.11$\\
\hline
$10^{-4}$
&$0.93\pm 0.04$
&$0.88\pm 0.04$
&$0.85\pm 0.10$
&$0.90\pm 0.07$\\
\hline
\end{tabular}
\caption{Ratio of apparent elastic moduli deduced from
  \eqref{eqn:wnalpha} for $n=1$ and $n=2$ to the ``global'' values.}
\label{tab:modul2}
\end{table}
This is further confirmed by another set of static response
calculations, in which displacements, rather than forces are
imposed. Let us define regularly spaced horizontal lines through the sample at $y=k
L_2/(n+1)$ for $0\le k \le n+1$ (so that $k=0$ corresponds to the
bottom and $k=n+1$ corresponds to the top), with $n$ an odd
number. Let us impose constant displacements $u=0$ in direction $\alpha$ on
line $k$ if $k$ is even, $u=(-1)^l u_0$ on line $k$ if $k=2l+1$ is odd. In
a homogeneous elastic system, the displacement field varies linearly
between neighboring lines $k$ and $k+1$, and the total elastic energy
reads
\be
\Check W_\alpha^{(n)}  = \frac{(n+1)^2 L_1}{2L_2}C_{\alpha} u_0^2
\label{eqn:w2}
\ee
Imposing to the center of each particle crossed by the $k$ lines
the displacements of the same point in a
homogeneous continuum (and leaving the rotations free), computing the
total elastic energy and using formula~\ref{eqn:w2}, one deduces other
values for $C_{22}$ and $C_{33}$, denoted as $\Check C_\alpha (n)$, given in
table~\ref{tab:modul3} for $n=1$ and $n=3$. Once again, they are
fairly close to the ``global'' moduli of table~\ref{tab:moduli}),
especially for the higher $p$ values, albeit slightly larger.
\begin{table}[htb]
\centering
\begin{tabular}{|c|cccc|}  %\cline{1-5}
\hline
$p$&$\Check C_{22}(1)/C_{22}$&$\Check C_{22}(3)/C_{22}$&$\Check C_{33}(1)/C_{33}$&$\Check C_{33}(3)/C_{33}$\\
\hline
$10^{-7}$
&$1.01\pm 0.02$
&$1.09\pm 0.03$
&$1.26\pm 0.16$
&$2.49\pm 0.69$\\
\hline
$10^{-6}$
&$1.01\pm 0.01$
&$1.07\pm 0.07$
&$1.28\pm 0.20$
&$2.05 \pm 0.45$\\
\hline
$10^{-5}$
&$1.01\pm 0.02$
&$1.06\pm 0.03$
&$1.15\pm 0.11$
&$1.65\pm 0.30$\\
\hline
$10^{-4}$
&$1.00\pm 0.01$
&$1.03\pm 0.02$
&$1.08\pm 0.05$
&$1.37\pm 0.14$\\
\hline
\end{tabular}
\caption{Ratio of apparent elastic moduli deduced from
  \eqref{eqn:w2} for $n=1$ and $n=3$ to the ``global'' values.}
\label{tab:modul3}
\end{table}
We therefore conclude that the elastic moduli are only weakly affected
by finite size effects.

\section{1D chain of identical balls}

We model the 1D chain of identical balls with $N$ identical particles
of mass $m$, separated by distance $\ell$, and connected by linear
springs of stiffness $K$. The first and last ball is also connected with
identical springs to walls. This system models small amplitude oscillations
of Hertzian balls under finite static pressure.  The $n$-th eigenmode of this
simple linear system is given by
\begin{equation}
u^{(n)}_x(t) = \sin(k_n x) \sin(\omega_n t + \phi_n) \,,
\end{equation}
where $u(t)$ is the displacement of a ball, and we label the balls by their
position $x=i\ell$, where $i=1,2,\dots,N$. The wavenumbers and
eigenfrequencies are
\begin{eqnarray}
\label{eq:dispersion}
k_n = \frac{n\pi}{(N+1)\ell}\,,\qquad
\omega_n = 2\sqrt\frac{K}{m} \sin\frac{k_n\ell}{2}\,.
\end{eqnarray}
The above dispersion relation determines the longitudinal sound speed
\begin{equation}
c=\left.\frac{d\omega}{dk}\right|_{k=0} = \ell\sqrt\frac{K}{m}
\end{equation}
by which the long wavelength waves propagate. The full solution is given by
\begin{equation}
\label{eq:soln}
u_x(t) = \sum_{n=1}^N a_n \sin(k_n x) \sin(\omega_n t + \phi_n)\,,
\end{equation}
where the amplitudes $a_n$ are computed by projecting the initial condition
onto the modes.  The two cases considered here are (A) $\dot u_1(t=0)=c$, and
(B) $u_1(t=0)=\ell$; all other displacements and velocities at $t=0$ are zero.
This gives
\begin{equation}
\label{eq:ic}
a^\text{A}_n = \frac{2\ell}{N+1} \cos\frac{k_n\ell}{2}\,,\qquad
a^\text{B}_n = \frac{2\ell}{N+1}  \sin k_n\ell\,.
\end{equation}
We are interested in the time dependence of the force the $N$-th ball presses
the wall:
\begin{equation}
\label{eq:force}
F(t) = Ku_N(t) \,.
\end{equation}
We cannot calculate this in closed form, but can for example look at its value
at the time ``a wave would arrive'': $t_0=N\ell/c$.
Substituting the first initial condition's $a^{A}$ into Eq.~(\ref{eq:soln})
and rewriting the highly oscillating terms we get
\begin{eqnarray}
F^A(t_0)=\sum_{n=1}^N && \frac{2K\ell}{N+1}\cos\frac{k_n\ell}{2}\sin{k_n\ell}
\\
\nonumber && \times \sin\left(k_n\ell + \frac{k_n^3}{k_0^3} +
\frac{O(k_n^5\ell^2)}{k_0^3}\right)
\end{eqnarray}
where $k_0=(24/N)^{1/3}/\ell$. The terms in the sum become highly oscillating
for $k_n\agt k_0$, effectively canceling each other.  The dominant
contribution therefore comes from terms with $k_n<k_0$ (or equivalently $k_n <
\text{constant}\times k_0$, as ultimately we will only compute scaling
exponents).  The sum of the slowly varying terms is approximated by integral:
\begin{eqnarray}
F^A(t_0)\sim\int_0^{k_0}&&\frac{2K\ell dk}{\pi}\cos\frac{k\ell}{2}\sin{k\ell}
\\ \nonumber
&& \times \sin\left(
k\ell + \frac{k^3}{k_0^3} + \frac{O(k^5\ell^2)}{k_0^3}\right)\,.
\end{eqnarray}
We look at the asymptotics $N\to\infty$ (implying $k_0\to 0$), for which we
have to find the terms lowest order in $k_0$. This yields
\begin{equation}
F^\text{A}(t_0)\sim k_0^2.
\end{equation}
Interestingly the second initial condition gives a different answer:
\begin{equation}
F^\text{B}(t_0)\sim k_0^3.
\end{equation}
The above two relations immediately lead to Eqs.~(\ref{eq:fA}--\ref{eq:fB}).

\bibliography{granularref}

\begin{thebibliography}{51}
\bibitem[{ult()}]{ultrasound}
\bibinfo{note}{Ultrasound is the word typically used for sound at frequencies
  higher than the upper frequency of the human ear, i.e. higher than about 20
  kHz. In the experiments to which we will compare, the typical frequency is of
  order 0.5 MHz.}
\end{thebibliography}

\end{document}